\documentclass[12pt,preprint]{aastex}

\slugcomment{accepted by ApJ 2 January 13 2012}

\shorttitle{The Red Flux of Brown Dwarfs}
\shortauthors{Leggett et al.}

\begin{document}

\title{The Properties of the 500~K Dwarf  UGPS J072227.51$-$054031.2,\\
 and a Study of the Far-Red Flux of Cold Brown Dwarfs. 
\footnote {Based on observations obtained at the Gemini Observatory, which is operated by the 
Association of Universities for Research in Astronomy, Inc., under a cooperative agreement 
with the NSF on behalf of the Gemini partnership: the National Science Foundation (United 
States), the Science and Technology Facilities Council (United Kingdom), the 
National Research Council (Canada), CONICYT (Chile), the Australian Research Council (Australia), 
Minist\'{e}rio da Ci\^{e}ncia e Tecnologia (Brazil) 
and Ministerio de Ciencia, Tecnolog\'{i}a e Innovaci\'{o}n Productiva (Argentina); also based on
data collected at Subaru Telescope, which is operated by the National Astronomical Observatory of Japan; and also based on observations made at the UK Infrared Telescope, which operated by the Joint Astronomy Centre on behalf of the Science and Technology Facilities Council of the UK.
 }
}

\author{S. K. Leggett\altaffilmark{1}}
\email{sleggett@gemini.edu}
\author{D. Saumon\altaffilmark{2}}
\author{M. S. Marley\altaffilmark{3}}
\author{K. Lodders\altaffilmark{4}}
\author{J. Canty\altaffilmark{5}}
\author{P. Lucas\altaffilmark{5}}
\author{R. L. Smart\altaffilmark{6}}
\author{C. G. Tinney\altaffilmark{7}}
\author{D. Homeier\altaffilmark{8}}
\author{F. Allard\altaffilmark{8}}
\author{Ben Burningham\altaffilmark{5}}
\author{A. Day-Jones\altaffilmark{9}}
\author{B. Fegley\altaffilmark{4}}
\author{Miki Ishii\altaffilmark{10}}
\author{H. R. A. Jones\altaffilmark{5}}
\author{F. Marocco\altaffilmark{6}}
\author{D. J. Pinfield\altaffilmark{5}}
\and
\author{M. Tamura\altaffilmark{11}}

\altaffiltext{1}{Gemini Observatory, Northern Operations Center, 670
  N. A'ohoku Place, Hilo, HI 96720} 
\altaffiltext{2}{Los Alamos National Laboratory, PO Box 1663, MS F663, Los Alamos, NM 87545, USA}
\altaffiltext{3}{NASA Ames Research Center, Mail Stop 245-3, Moffett Field, CA 94035, USA}
\altaffiltext{4}{Planetary Chemistry Laboratory, Department of Earth and Planetary Sciences, McDonnell Center for the Space Sciences, Washington University, St. Louis, MO 63130-4899, USA}
\altaffiltext{5}{Centre for Astrophysics Research, Science and Technology Research Institute, University of Hertfordshire, Hatfield AL10 9AB, UK}
\altaffiltext{6}{INAF/Osservatorio Astronomico di Torino, Strada Osservatorio 20, 10025 Pino Torinese, Italy} 
\altaffiltext{7}{Department of Astrophysics, University of New South Wales, NSW 2052, Australia}
\altaffiltext{8}{CRAL, Universit\'{e} de Lyon, \'{E}cole Normale Sup\'erieur, 46 all\'{e}e d'Italie, 69364 Lyon Cedex 07, France}
\altaffiltext{9}{Universidad de Chile, Camino el Observatorio No. 1515, Santiago, Chile, Casilla 36-D}
\altaffiltext{10}{Subaru Telescope, National Astronomical Observatory of Japan, 650 N. A'ohoku Place, Hilo, HI 96720, USA}
\altaffiltext{11}{National Astronomical Observatory of Japan, 2-21-1 Osawa, Mitaka, Tokyo 181-8588, Japan}

\begin{abstract}
We present $i$ and $z$ photometry for 25 T dwarfs and one L dwarf.
  Combined with published photometry, the data show that the $i - z$, $z - Y$ and $z - J$ colors of T dwarfs are very red, 
and continue to increase through to the late-type T dwarfs, with a hint of a saturation for the latest types
with $T_{\rm eff} \approx 600$~K.
We present new 0.7 -- 1.0~$\mu$m and 2.8 -- 4.2~$\mu$m spectra for the very late-type T dwarf
 UGPS J072227.51$-$054031.2, as well as improved astrometry for this dwarf. Examination of the spectral energy distribution using the new and published data, with Saumon \& Marley models, shows that the dwarf has $T_{\rm eff} = 505 \pm 10$~K,  a mass of 3 -- 11~$M_{\rm Jupiter}$ and an age between 60~Myr and 1 Gyr. This young age is consistent with the thin disk kinematics of the dwarf. 
The mass range overlaps with that usually considered to be
planetary, despite this  being an unbound object discovered
in the field near the Sun.
This apparently young rapid rotator is also undergoing vigorous atmospheric mixing, as determined by the IRAC and WISE-2 4.5~$\mu$m photometry and the  Saumon \& Marley models. The optical spectrum for this 500~K object shows 
clearly detected lines
of the neutral alkalis Cs and Rb, which  are emitted from deep atmospheric layers with temperatures of  900 -- 1200~K. 
\end{abstract}

\keywords{stars: brown dwarfs, stars: abundances, line: profiles}

\section{Introduction}

Cool field brown dwarfs (i.e. stellar-like objects with a mass below that 
required for the onset of hydrogen fusion, see e.g. Hayashi \& Nakano 1963, Kumar 1963, Burrows \& Liebert 1993) are interesting for many reasons. They are important for studies of star formation and the initial mass function at low masses, as well as for studies of the chemistry and physics of cool atmospheres. Furthermore, they can act as proxies for giant exoplanets which have similar temperatures and radii, but which have less well-understood interiors and which are harder to observe -- brown dwarf science, for example, is directly applicable to the HR 8799 planets (Marois et al. 2008). 
The far-red Sloan Digital Sky Survey (SDSS; York et al. 2000), the near-infrared Two Micron All Sky Survey (2MASS; Skrutskie et al. 2006) and the UKIRT Infrared Deep Sky Survey (UKIDSS; Lawrence et al. 2007) led to the discovery of significant numbers of very low-mass and low-temperature stars and brown dwarfs in the field -- the L and T dwarfs (e.g. Kirkpatrick et al. 1997, Strauss et al. 1999, Leggett et al. 2000, Warren et al. 2007). The L dwarfs have effective temperatures ($T_{\rm eff}$) of 1400 -- 2200~K and the T dwarfs have $T_{\rm eff} <$ 1400~K (e.g. Stephens et al. 2009). 

Until very recently, the coolest brown dwarfs known were six T9 -- T10 dwarfs with $T_{\rm eff} = 500$--600~K  (Warren et al. 2007; Burningham et al. 2008; Delorme et al. 2008, 2010; Lucas et al. 2010; Mainzer et al. 2011).  These objects have masses of 10 -- 20~$M_{\rm Jupiter}$ (e.g. Leggett et al. 2010a; however here we determine a mass $\leq 11$~$M_{\rm Jupiter}$ for one of these dwarfs, see \S 3). In 2011 two very faint companions were found that appear to have $T_{\rm eff}$ values between 300~K and 400~K; one is a close companion to a T9.5 dwarf (Liu et al. 2011), and the other a more distant companion to a white dwarf (Luhman, Burgasser \& Bochanski 2011). 
Also, Cushing et al. (2011) have announced the discovery by 
the Wide-field Infrared Survey Explorer mission (WISE; Wright et al. 2010) of six isolated objects with extremely red [3.4]-[4.6] (W1-W2) colors. Followup near-infrared spectroscopy by the WISE team shows
that their near-infrared absorption bands are broader
than for T9 dwarfs, the dwarfs increase dramatically in brightness from the near-infrared to the 
mid-infrared, and they appear to be intrinsically very faint 
at near-infrared wavelengths.
For these reasons Cushing et al. propose that five of these six
new brown dwarfs be classified as Y0, with one dwarf being classified
as even later.
 Cushing et al. estimate temperatures of $<$ 300~K to 450~K for the six dwarfs, by fitting synthetic spectra to the near-infrared spectra (but see \S 3.1).  Assuming an age range of 1 -- 10 Gyr, these isolated solar neighbourhood objects have masses in the range 5 -- 20~$M_{\rm Jupiter}$.

Significant chemical changes occur in a brown dwarf atmosphere as it cools (e.g. Kirkpatrick et al. 1999; Lodders 1999; Geballe et al. 2002; Knapp et al. 2004; Cushing et al. 2006, 2011). Cloud decks of liquid and solid iron and silicate grains exist in the photospheres of L dwarfs, and their grain-reddened spectral energy distributions show strong metal hydride and neutral alkali lines, as well as H$_2$O, CO and, in the mid-infrared, CH$_4$ bands. The T dwarf photospheres are too cool for iron and silicate grains, which condense in deeper layers of the atmosphere and have little to no 
role in forming the spectrum emitted by the photosphere. Note, however, that
in the atmospheres of late-type T dwarfs, the alkalis start condensing into salts and $\rm Na_2S$ solids (\S 4.1), and
 Burgasser et al. (2011) find that condensate clouds may be required to reproduce the near-infrared spectra of young, low gravity, T dwarfs.
The predominantly clear T dwarf photospheres have optical flux distributions shaped  by
the red wing of the strong K I doublet at 770~nm, while their near- 
and mid-infrared spectra
are sculpted by strong bands of  H$_2$O, CH$_4$ and, in the mid-infrared, NH$_3$. Recently, Bochanski et al. (2011) have shown that 
NH$_3$ is also detectable in high-resolution near-infrared spectra of the $T_{\rm eff} \approx $500~K very late-type T dwarf UGPS J072227.51$-$054031.2 (hereafter UGPS 0722$-$05, Lucas et al. 2010).
At the low temperatures of T dwarf photospheres, alkalis are turning into chloride gases and the neutral alkalis are starting to disappear. At even lower temperatures, alkali chloride gases condense out as salts. For details on alkali chemistry see Lodders (1999).

For a $T_{\rm eff} \approx $1000~K T7 brown dwarf   we expect that: the refractory elements Mg, Al, Ca and Ti will all have been removed from the photosphere; carbon will be primarily in the form of CH$_4$; nitrogen will be primarily in the form of N$_2$; and the Cs, K, Na and Rb neutral alkalis will be disappearing from the photosphere as they  turn into chloride gases. For T8 -- T10 dwarfs with $800 \geq T_{\rm eff}$~K $\geq 500$, 
we expect that nitrogen will be primarily in the form of NH$_3$ (depending on atmospheric mixing, see \S 3.5) and
the neutral alkali lines will be weak. 
This interpretation is over-simplified however,
because the background opacity (the pseudo-continuum) must be accounted for, and the temperature where the feature is formed can be very different from the $T_{\rm eff}$ value. 
The red spectral region of a brown dwarf contains strong absorption lines of Cs, K, Na and Rb, and is the best region for studying the evolution of the alkali features with decreasing  $T_{\rm eff}$ (e.g. Burgasser et al. 2003).
A brown dwarf atmosphere is quite transparent in the far-red, meaning that the emergent flux is formed in a relatively hot zone where neutral alkalis can still exist. 

Understanding the alkali chemistry of brown dwarfs is not only of intrinsic interest, it is important for modelling their red colors for the interpretation of sky surveys,
see the discussion in Marley et al. (2002). Sky surveys sensitive to the red or far-red colors of brown dwarfs include:
the CFHT Brown Dwarf Survey (DeLorme et al. 2008), the UKIRT Infrared Deep Sky Survey (UKIDSS; Lawrence et al. 2007), Pan-Starrs (Kaiser 2004), the Visible and Infrared Survey Telescope for Astronomy (VISTA; Emerson 2001), and the Large Synoptic Survey Telescope (LSST; Ivezic et al. 2011). Getting the alkali chemistry right is also critical for understanding exoplanets. When planets are imaged in reflected light, alkalis can shape the spectra of the warmer planets that do not have water clouds or other hazes.  Some irradiated planets are modelled to have strong Na and K features (Sudarsky et al. 2000, Cahoy et al. 2010), and  Na and K absorption has been detected in the spectra of extrasolar planet atmospheres (e.g. Charbonneau et al. 2002; Jensen et al. 2011; Redfield et al. 2008; Sing et al. 2008, 2011; Snellen et al. 2008).

In \S 2 of this paper we present new optical spectra for two brown dwarfs with $T_{\rm eff} \approx 500$~K and 750~K, as well as new $iz$ photometry for a sample of T dwarfs. We present additional data for the 500~K dwarf UGPS 0722$-$05: a 2.8 -- 4.2~$\mu$m spectrum and improved astrometry. We use these data, together with other published data and the models of Saumon \& Marley (2008) and Saumon et al. (2012), to perform a detailed analysis of the spectral energy distribution of this brown dwarf, in \S 3. Having determined the atmospheric properties of UGPS 0722$-$05, we use the optical spectrum for this dwarf to examine the strength of the neutral alkalis at cold temperatures, in \S 4. Also in \S 4, we compare the far-red to near-infrared colors of the latest-type T dwarfs to the Saumon \& Marley  models. For objects not much cooler than  UGPS 0722$-$05 we can expect to see pronounced changes in the optical colors of the coolest brown dwarfs, as larger fractions of the alkalis are converted into chloride gases and condensation of alkali chloride salts begins. Finally, in \S 5, we present trends in $izYJH$ with spectral type, and discuss the detectability of cold brown dwarfs by far-red/near-infrared ground-based surveys. Our conclusions are given in \S 6.

\section{Observations}

\subsection{Optical Spectra of 2MASS 0415$-$09 and UGPS 0722$-$05}

Optical spectra were obtained with the  Gemini Multi-Object Spectrograph (GMOS, Hook et al. 2004)
of the T8 dwarf 2MASSI J0415195$-$093506 (hereafter 2MASS 0415$-$09; Burgasser et al. 2002a) and of the T10 dwarf  UGPS 0722$-$05 (Lucas et al. 2010). These objects were chosen as they are cool, while covering a range in $T_{\rm eff}$ and being bright enough to make the observation feasible on an 8-m telescope. 2MASS 0415$-$09 has $T_{\rm eff} = 750$~K (Saumon et al. 2007) and UGPS 0722$-$05 has $T_{\rm eff} \approx 500$~K (Lucas et al. 2010,  \S 3).

GMOS at Gemini North was used, through queue time granted under program GN-2010B-Q-59. The
R400 grating was used with the OG515 blocking filter.  The central wavelength was 840~nm, with wavelength coverage of 
620 -- 1010 nm. Detector fringing affected the spectra longwards of about 800 nm, which was mitigated by coadding 
spatially-offset sky-subtracted spectral images.
The 0$\farcs$5 slit was used with $2\times 2$ binning, and the resulting resolution was  3.5 \AA\
 as determined from the arc images.
An observing log is given in Table 1.

Flatfielding and wavelength 
calibration were achieved using the quartz halogen and CuAr lamps in the on-telescope calibration unit. 
The spectrophotometric standard EG 131 was used to determine the instrument response curve, 
and to flux calibrate the spectra. The data were reduced using routines 
supplied in the IRAF Gemini package.
Figure 1 shows the GMOS spectra; the signal to noise ratio is typically around 10 at the fainter blue end of each dwarf's spectrum, and around 20 or 30
at the brighter red end for the 2MASS 0415$-$09 spectrum and the UGPS 0722$-$05 spectrum, respectively.

\subsection{$i$ and $z$ Photometry for a Sample of T dwarfs}

Photometry was obtained for a sample of T dwarfs using the GMOS instruments at both Gemini North and South,
through queue time granted under programs GN-2010A-Q-81 and GS-2010B-Q-39. Objects were chosen 
that were bright enough to make $i$-band photometry (where there is very little flux) feasible on an 8-m telescope, and that provided (when combined with published data)
adequate sampling of the spectral type versus $i - z$ relationship. Photometry was also obtained for one L dwarf companion to a T dwarf, and for T dwarfs with SDSS photometry,  in order to determine the transformation between the GMOS and SDSS systems. A total of 25 T dwarfs and one L dwarf was observed. 

The sample is listed in the observing log given in Table 2.
Exposure times are also given in Table 2; the targets were observed in a multi-position offset pattern and fringe frames for subtraction were created for each target in each filter using these multiple images. Twilight flats were used for flatfielding, and Landolt (1992) or SDSS photometric standard fields (Smith et al. 2002, \url{http://www-star.fnal.gov/Southern\_ugriz/www/Fieldindex.html}) were used for calibration. The extinction coefficients for each filter at each site given on the Gemini GMOS calibration web page were adopted: for Gemini North 0.10 and 0.05, for Gemini South 0.08 and 0.05, magnitudes airmass$^{-1}$ at $i$ and $z$ respectively. Color terms are small and were not used. Photometry was derived using apertures with radii 6 or 10 pixels, or diameters 1$\farcs$7 and 2$\farcs$8. 
The data were reduced using routines 
supplied in the IRAF Gemini package.

Table 2 gives the derived $i$ and $z$ photometry in the GMOS-North and GMOS-South natural systems (see \S 2.2.1).

\subsubsection{$iz$ Photometric Systems}

\subsubsubsection{T dwarfs and Photometric Systems}

The extremely structured nature of the energy distribution of T dwarfs leads to large differences in the photometry obtained with different photometric systems. Stephens \& Leggett (2004) demonstrate this for near-infrared filter sets. In the far-red the very rapidly rising flux to the red, combined with the fact that the red edge of the $z$ bandpass is usually defined by the detector cut-off, add significant complications. Spectra for the very late-type T dwarfs 2MASS 0415$-$09 and UGPS 0722$-$05 are shown in Figure 1, along with filter bandpasses.

We have synthesized $i$ and $z$ photometry for a sample of L and T dwarfs using flux-calibrated red and infrared spectra that span the filter bandpasses. We find that the synthesized SDSS- and GMOS-system values differ from the measured values by 10 -- 20\%,  which can be accounted for by small discrepancies in the red cut-offs of the filters.

\subsubsubsection{GMOS-North and GMOS-South}

The GMOS North and South systems are very similar -- the filters are designed to be identical, but the E2V detector responses are slightly different (see Figure 1). Because of the difficulty in synthesizing the photometry, we use the repeat measurements of three brown dwarfs to determine the offset between the two systems. Table 2 shows that, for the T0, T4.5 and T8 dwarfs observed at both sites, the  values obtained in the South are $0.05 \pm 0.04$ magnitudes fainter than in the North. We convert the South to the North system by subtracting 0.05 magnitudes from the $i$ and $z$ values measured with GMOS-South. 

\subsubsubsection{GMOS-North and SDSS}

There are significant differences between the GMOS and SDSS $i$ and $z$ bandpasses (see Figure 1). Differences at the red end of each filter will be especially important due to the rapidly increasing flux to the red. Of the twenty-six objects in Table 1, eleven have SDSS (Data Release 8, DR8; Aihara et al. 2011) $i$-band measurements and thirteen have SDSS $z$-band measurements. Brown dwarfs with  SDSS data were deliberately observed in order to determine the photometric transformations.

Figure 2 plots the difference between the SDSS and GMOS-North $i$ and $z$ values, as a function of $i - z$ (as measured on GMOS North). Weighted (by the inverse of the square of the uncertainty) fits give:
$$ i_{SDSS} - i_{GMOS} = 0.723 - 0.0521\times(i - z)_{GMOS} $$
$$ z_{SDSS} - z_{GMOS} = 0.221 - 0.0712\times(i - z)_{GMOS} $$
for $ 1.9 < (i - z)_{GMOS} < 4.0$.
Similarly, Figure 2 also shows $\delta z$ as a function of $z_{GMOS} - J_{MKO}$. A weighted fit gives:
$$ z_{SDSS} - z_{GMOS} = 0.318 - 0.0924\times(z_{GMOS} - J_{MKO}) $$
for $ 2.7 < z_{GMOS} - J_{MKO} < 4.3$. Omitting four $i$ band datapoints with $\sigma (\delta i) \geq 0.5$, the $rms$ scatter around the $ i_{SDSS} - i_{GMOS}$ fit is 0.06 magnitudes. Omitting four $z$ band datapoints with $\sigma (\delta z) \geq 0.1$, the $rms$ scatter around the $ z_{SDSS} - z_{GMOS}$ fit, as a function of $i - z$, is 0.08 magnitudes, and as a function of $z - J$ it is 0.10 magnitudes. Tests of the transformations for $z$ using both  $i - z$ and  $z - J$ show that the difference in the derived $ z_{SDSS}$ is around 0.03 magnitudes.
We adopt an estimated uncertainty in the GMOS-North to SDSS system transformations of 8\%.

Table 3 lists the transformed GMOS photometry, together with SDSS DR8 photometry where available (note that there can be significant differences in the photometry between SDSS Data Releases and it is advisable to use the latest release). An 8\% error has been added in quadrature to the measurement error to allow for the uncertainty in the GMOS to SDSS transformation, and an additional 4\%  is added in quadrature for measurements made using GMOS-South. The $z - J$ relationship has been used to transform the $z$ magnitudes as opposed to the $i - z$ relationship, due to the smaller uncertainties in the $J$ magnitudes. The difference between the $z$ values determined from the two transformations is small: $0.01 \pm 0.03$ magnitudes.

\subsubsubsection{Other $iz$ Systems}

Additional $z$-band photometry is available for 52 T dwarfs identified in the UKIDSS database. These data were obtained using   the European Southern Observatory (ESO) Multi-Mode Instrument (EMMI) mounted on the New Technology Telescope (NTT) at La Silla, Chile, and are published in Warren et al. (2007), Lodieu et al. (2007), Pinfield et al. (2008), Burningham et al. (2008, 2010). Warren et al. use spectrophotometry for a set of T dwarfs to calculate $ z_{SDSS} = z_{EMMI} + 0.2 $, and this offset has been applied in the UKIDSS publications.

$z$-band photometry is also available for three T dwarfs identified in the CFBDS dataset. These data were obtained using MegaCam on the CFHT, and are published in Delorme et al. (2008, 2010).  Reyle et al. (2010) synthesize $i$ and $z$ photometry for L and T dwarfs using flux-calibrated spectra. Comparison of these values to measured SDSS values suggests a large difference between the systems of $\sim 0.3$ magnitudes, where the MegaCam values are fainter. We have synthesized CFHT and SDSS $z$ magnitudes using the files available for the MegaCam optics, filters, and detector, as well as the telescope optics, available from the CFHT web pages 
\footnote{at \url{http://www.cfht.hawaii.edu/Instruments/Imaging/MegaPrime/data.MegaPrime/}: MegaCam\_Filters\_data.txt, CFHT\_Primary\_Transmission.txt, CFHT\_MegaPrime\_Transmission.txt, E2V\_CCD42-90\_QEmodel.txt.},
and find a smaller difference. Synthesizing SDSS and MegaCam values indicates that $ z_{SDSS} = z_{MegaCam} - C$ where $C = 0.1$ for early-type T dwarfs and  $C = 0.2$ for late-type T dwarfs. We have applied this smaller correction to the photometry for the three CFBDS dwarfs.

\subsection{2.8 -- 4.2 $\mu$m Spectroscopy of UGPS 0722$-$05}

A 2.8 -- 4.2~$\mu$m spectrum was obtained for UGPS 0722$-$05 using the Infrared Camera and
Spectrograph (IRCS, Kobayashi et al. 2000) on the Subaru telescope. 
UGPS~0722-05 was observed  on the nights of 23 and 24
January 2011 with the $L$-band grism, using a $0\farcs9$ slit; the spectral
resolution is R $\approx$ 180. The total integration time was 120 minutes,
made up of 48 minutes from 23 January and 72 minutes from 24 January.
Observations were made in an ABBA pattern with 60~s integrations and a
$6\arcsec$ nod. 

The
spectrum was extracted in IRAF using a $0\farcs75$ aperture. Since no
suitable arc lamp is available a linear solution was adopted for the
wavelength calibration, based on two widely spaced telluric features,
following the procedure described in the IRCS data reduction cookbook 
(\url{http://www.subarutelescope.org/Observing/DataReduction/index.html}).
The estimated accuracy in the wavelength calibration is $\sim$1~nm. 

Two A- -- F-type stars were observed
as telluric standards on each night, before and after the observation
of UGPS~0722-05.  The extracted telluric-calibrated spectra
from 23 January and 24 January were found to be very
similar, with the exception of a slight disagreement in the slope of the
continuum at the longest wavelengths (4.00 -- 4.18~$\mu$m). The spectrum 
was flux calibrated using the IRAC [3.6] photometry,
which has an uncertainty of 5\% (\S 3.2, Table 4).

Figure 3 shows the final reduced spectrum.  The error spectrum is also shown, 
which is based on the difference between the two subsets of data. The signal to noise ratio varies across the spectrum, and is around 5 to 15 from 3.7~$\mu$m to 4.1~$\mu$m,
the wavelength region less impacted by strong  CH$_4$ absorption.
The spectrum has 
significant structure around 3.8~$\mu$m: there are absorption features at 
3.71 -- 3.74~$\mu$m, 3.77 -- 3.79~$\mu$m and 3.83 -- 3.86~$\mu$m. The Saumon \& Marley models
show that all the structure in this region is due to absorption by CH$_4$, 
and the models reproduce the observations quite well (\S 3.4). These models show 
the absorption bands around 3.8~$\mu$m strengthening with decreasing temperature,
with the features at 3.72~$\mu$m and 3.84~$\mu$m only being apparent for $T_{\rm eff}
\lesssim 800$~K, depending on resolution. There is possibly a hint of the structure in the 
R $\approx$ 600 spectrum of the 900 -- 1000~K dwarf $\epsilon$ Indi Bb presented by King et al. (2010), but the  R $\approx$ 100 AKARI spectrum of the 750~K dwarf 2MASS 0415$-$09
presented by Yamamura, Tsuji \& Tanab\'{e} (2010)
is too coarse to show these features.

\subsection{An Improved Trigonometric Parallax for  UGPS 0722$-$05}

In Lucas et al. (2010)  we found a parallax for UGPS 0722$-$05 of 237
$\pm$ 41 mas based on seven UKIRT observations from 2006 -- 2010; adding a very
low signal-to-noise 2MASS detection gave a value of 246 $\pm$ 33 mas. The
UKIRT observations for this object have continued, and  we now have 
23 observations covering a period from November 2006 --
April 2011. Following the procedures for observing, image treatment and
astrometric reduction described in Marocco et al. (2010), we determine
parallax and proper motions that are very similar in value to those presented in Lucas et al. 
but with much smaller, $< 1$~\%, errors.
The revised values are given in Table 4.

Incorporating the 2MASS observation did not improve the precision of our result.
Also, the UKIRT observations provide fainter reference objects therefore
the correction from relative to absolute parallax is smaller and more
accurate. The high precision found for this object is consistent with
that found in Marocco et al. (2010) for targets of similar brightness
 with similar temporal baselines.

\section{Analysis of the Physical Properties of UGPS 0722$-$05}

\subsection{Estimates of  $T_{\rm eff}$, Gravity, Mass and Age}

In this Section we perform an analysis of the spectral energy distribution 
of UGPS 0722$-$05
in order to constrain $T_{\rm eff}$, gravity ($g$), mass and age for the object.  Lucas et al. (2010) classify the dwarf as T10, and find $T_{\rm eff} = 520 \pm 40$~K based on its luminosity. Cushing et al. (2011) reclassify the dwarf as T9, and estimate $T_{\rm eff} = 650$~K based on a fit of synthetic spectra to the near-infrared spectrum. Bochanski et al. (2011) present high-resolution near-infrared spectra for  UGPS 0722$-$05, and find that  synthetic near-infrared spectra indicate  
$T_{\rm eff} = 500$ -- 600~K (a 100~K warmer temperature is found if the  luminosity constraint is neglected),
although the fit is poor.   
Lucas et al. and Bochanski et al. use the BT-SETTL models of Allard et al. (2007a),
while Cushing et al. (and this work) use the models of  Saumon \& Marley (2008).
Note that at these low temperatures the near-infrared spectrum makes up $<$ 30 \% of the total flux from the brown dwarf. 
Also, the opacity line lists for important molecules are known to be 
incomplete in the near-infrared and thus all models are subject to
systematic errors.
Hence the temperatures found by Bochanski et al. and  Cushing et al., based on near-infrared data alone, are prone to potentially significant error. In fact, Cushing et al. find that the near-infrared spectroscopic temperatures
imply distances to the dwarfs that can be very different from photometric or trigonometric parallax solutions.   
In this work we incorporate mid-infrared data and a new parallax  to reduce the uncertainty in the derived properties of the brown dwarf. 

The difference in spectral type assigned by Lucas et al. and Cushing et al. of T10 {\it cf.} T9 is important to note. A definitive spectral type determination for UGPS 0722$-$05 is beyond the scope of this paper. The T10 classification is based on a comparison of the dwarf's spectral indices to those of the T8 and T9 dwarfs, where the T9 dwarfs are classified based on the extension of the Burgasser et al. (2006) scheme by Burningham et al. (2008).  However Cushing et al. propose that 
 UGPS 0722$-$05 is T9 based on a comparison of 1.1 -- 1.7~$\mu$m spectra for the T6, T7 and T8 spectral standards of Burgasser et al. (2006). The definition of the end of the T sequence and the start of the Y sequence is 
likely to change as more dwarfs are identified with $300 \leq 
T_{\rm eff}$~K $\leq 600$; in the meantime we adopt a spectral type of T10 for  UGPS 0722$-$05, as found by Lucas et al..

\subsection{Observational Data and Kinematics}

In this analysis we use the near-infrared spectrum and near- and mid-infrared photometry given by Lucas et al. (2010), together with the new optical and $L$-band spectra presented here. To these data we add the WISE  mid-infrared photometry given by Kirkpatrick et al. (2011) for UGPS 0722$-$05. The optical, near-infrared and $L$-band spectra are flux-calibrated using the GMOS $z$ photometry presented here, the $JHK$ photometry presented in Lucas et al., and the  IRAC [3.6] photometry presented in Lucas et al., respectively.

We also use the new values for the trigonometric parallax and proper motions presented here.  We combine these data with the radial velocity determined by Bochanski et al. (2011), to determine $UVW$ velocities for UGPS 0722$-$05. We find that the kinematics of this brown dwarf are typical of thin disk objects, in agreement with Bochanski et al. (see e.g. Figure 21 of Kilic et al. 2010, which is based on thick disk and halo velocities from Chiba \& Beers 2000, and thin disk kinematics from Soubiran, Bienayme \& Siebert 2003). Bochanski et al. find that  UGPS 0722$-$05 has a rotational velocity of $40 \pm 10$~km\,s$^{-1}$, which is rapid, but similar to the values found for other T dwarfs (e.g. Zapatero Osorio et al. 2006). Note that the velocity is computed by comparison to a model spectrum and is thus 
sensitive to the treatment of line broadening.

Table 4 lists the available observational data for  UGPS 0722$-$05.

\subsection{Luminosity-Implied Parameters}

We compare the optical to 4~$\mu$m spectrum of UGPS 0722$-$05 to solar-metallicity cloudless models described in Saumon \& Marley (2008),
with updated H$_2$ collision-induced absorption and the new NH$_3$ line
list of Yurchenko, Barber \& Tennyson (2011). These new models are described in Saumon et al. (2012). Currently, only solar-metallicity models are available for this new grid, however the photometric colors of  UGPS 0722$-$05 imply that it has a metallicity close to solar (Lucas et al. 2010) and brown dwarf evolution is insensitive to small deviations from solar  metallicity  (e.g. Saumon \& Marley 2008, Leggett et al. 2010b, Burrows et al. 2011).

The inclusion of the 4~$\mu$m flux peak is useful for covering the spectral energy distribution, and we find that the spectral data (the far-red spectrum and the 3 -- 4 ~$\mu$m spectrum presented here, together with the near-infrared spectrum of Lucas et al. 2010) sample 30 -- 40 \% of the derived bolometric flux (see below).
For each model, defined by $T_{\rm eff}$ and gravity, there is an associated radius determined by evolutionary models (Saumon \& Marley 2008). The analysis follows the self-consistent luminosity method described in Leggett et al. (2010b), and Saumon et al. (2006, 2007). Briefly, the observed spectrum is integrated to give an observed flux ($5.17 \pm 0.13 \times 10^{-13}$ erg\,s$^{-1}$cm$^{-2}$), and
synthetic spectra are used to determine a bolometric correction. A family of ($T_{\rm eff}$, $g$) values are derived that provide bolometric luminosity ($L_{\rm bol}$) values consistent with both the synthesized bolometric correction and the luminosity implied by evolutionary models.   The $L_{\rm bol}$ values vary slightly due to the differences between the models used for the unmeasured parts of the spectral energy distribution.
We limit the allowed age range for the dwarf to between 40 Myr (defined by assuming it will be more massive than 2~$M_{\rm Jupiter}$) and 7~Gyr (the thin disk kinematics suggest the dwarf is younger than 10 Gyr). We show below in \S 3.5 that the warmer effective temperatures 
associated with ages greater than 7~Gyr do not fit the mid-infrared
photometry and are less likely.

Figure 4 shows a plot of $T_{\rm eff}$ against log $g$ and illustrates the  solutions that we obtain with our luminosity method, superimposed on the cloudless cooling tracks of Saumon \& Marley (2008). 
 Table 5 lists three representative solutions
  (shown as solid dots in Figure 4) for $T_{\rm eff}$, log $g$, log $L_{\rm bol}$, mass and radius, which span the
  adopted range in age.   The uncertainty in log $L_{\rm bol}$ is dominated by the uncertainty in the  flux calibration of the spectrum (2.6\% overall), and  is  $\pm$0.014 dex, or  $\pm$4~K in
 $T_{\rm eff}$, for a fixed gravity. These are formal uncertainties based
on those for the flux calibration and the parallax and do not take into
account systematic biases in the models.
Our luminosity range of $6.8\times 10^{-7}$ to $8.9\times 10^{-7}$ $L_{\odot}$ is consistent with the 
lower range of the values found by Lucas et al. (determined by combining the near-infrared spectrum with mid-infrared photometry).

\subsection{Spectral Comparison to the Models}

Figure 5 shows the observed optical through near-infrared and $L$-band spectrum, and the synthetic spectra for the three  Saumon \& Marley models of Table 5. 
These model spectra include departures from chemical
equilibrium
caused by vertical mixing, parametrized with an eddy diffusion
coefficient of
$K_{zz} = 10^{5.5}$ cm$^2$ s$^{-1}$. At these wavelengths mixing does not significantly impact the spectra, however we find in \S 3.5 that  rapid mixing is needed to reproduce the 4.5~$\mu$m photometry.
In Figure 5 the synthetic spectra have not been scaled to fit, but are calibrated by the measured distance to the object and the radius implied by evolutionary models for each ($T_{\rm eff}$, $g$) value. None of the models fit the 1.0 -- 1.3~$\mu$m region well, although the highest temperature and highest gravity model (with less flux at 1.0 --- 1.3~$\mu$m) would be favored, with $T_{\rm eff} = 550$~K. 

In previous work we have found, as we do here, that the models overpredict the flux at the 1.0~$\mu$m $Y$-band and the 1.25~$\mu$m $J$-band, for brown dwarfs cooler than $\sim$ 700~K (e.g. Leggett et al. 2009, 2010a,b). It is known that the CH$_4$ opacity line list at these wavelengths is incomplete, and this is a possible explanation for the discrepancies seen in Figure 5 in the $Y$ and $J$ bands, although thin clouds may also affect the relative peak fluxes (e.g. Burgasser et al. 2011).

The models fit the 2.8 -- 4.2~$\mu$m spectrum reasonably well, although they are too faint at 4.1~$\mu$m,
which is probably due to the uncertain
CH$_4$ absorption that still dominates in these models at 4.1~$\mu$m. Note that
the models include the effects of vertical mixing on the CO abundance,
which has a strong band for $\lambda > 4.5\,\mu$m. The abundance of
CO$_2$, which has a strong band centered at 4.23$\,\mu$m, is also increased
by vertical transport (Burningham et al. 2011) but this effect is not
included here. An excess of CO$_2$ would further depress the
modeled flux in the 4.2 -- 4.4$\,\mu$m region. The three model spectra
  are very similar in the $L$-band wavelength region and do not constrain our fit.

\subsection{Mid-Infrared Photometric Comparison to the Models}

For brown dwarfs as cold as UGPS 0722-05, most of the flux
is emitted in the mid-infrared and the models can be usefully constrained
by the available photometry at 3 -- 12$\,\mu$m. 
Figure 6 shows a comparison of the synthetic photometry to the observed IRAC [3.6] and [4.5], $L^{\prime}$ and $N$ magnitudes given by Lucas et al. (2010), and Figure 7 shows the comparison for the WISE  W1(3.4), W2(4.6), W3(12) and W4(22)  values given by Kirkpatrick et al. (2011). The apparent magnitudes are plotted as a function of the vertical mixing coefficient $K_{zz}$ cm$^2$ s$^{-1}$. 
Generally, larger values of $K_{zz}$ imply greater enhancement of CO  over CH$_4$. Values of log $K_{zz} = 2$ -- 6,  corresponding to mixing timescales of $\sim 10$ yr to $\sim 1$ hr, respectively, reproduce the observations of T dwarfs (e.g. Saumon et al. 2007).
Figures 6 and 7 show that the mixing strongly impacts the IRAC  [4.5] and WISE W2 4.2 -- 5.0~$\mu$m  flux; rapid mixing
 enhances CO and CO$_2$ in this region, reducing the flux from the dwarf. These 
two datapoints constrain the mixing coefficient for UGPS 0722-05 to log $K_{zz} \approx 5.5$ -- 6.0. (The choice of mixing coefficient does not impact the physical parameters derived using our self-consistent luminosity method, see e.g. Leggett et al. 2010b).

The 8 -- 16~$\mu$m flux ($N$,  W3; Figures 6 and 7) constrains $T_{\rm eff}$ for
UGPS 0722-05 to be between 492~K and 518~K; the warmest 550~K model is excluded.
We find that a mixing coefficient of log $K_{zz} \approx 5.5$ -- 6.0 and $T_{\rm eff} \approx$  505~K  fits all the photometry except W1(3.4) and [3.6] (the $L^{\prime}$(3.8~$\mu$m) photometry is too uncertain to constrain the models).  W1 and [3.6]  effectively measure the same part of the spectrum, and Figure 5 shows that the  Saumon \& Marley models underestimate the flux at 3.5 -- 4.2~$\mu$m.  The discrepancy at 4.1~$\mu$m in particular has a strong impact on the photometry, given that this is the only region with significant flux within these particular passbands.
Figures 6 and 7 show that the synthetic photometry is too faint by $\sim$ 20 \% in this region.

Although no mid-infrared spectra for UGPS 0722$-$05 exist, these models have been validated using spectral data  for brown dwarfs with a similar $T_{\rm eff}$. The models reproduce the 7.5 -- 14.2~$\mu$m
{\it Spitzer} spectra well for ULAS J003402.77–005206.7 and ULAS J133553.45+113005.2 with  $T_{\rm eff} = 550$ -- 600~K and 500 -- 550~K respectively (Leggett et al. 2009), as well as for Wolf 940 B with  $T_{\rm eff} = 585$ -- 625~K (Leggett et al. 2010b).
Thus Figures 6 and 7 demonstrate that the models reproduce the spectral energy distribution at
$4.2 \lesssim \lambda \, (\mu$m$)\lesssim 16$ 
well, and that  
our  $L_{\rm bol}$ determination is robust.

\subsection{Adopted Parameters}

The 8 -- 16~$\mu$m flux ($N$,  W3; Figures 6 and 7) indicates that  UGPS 0722$-$05 has $T_{\rm eff}$ mid-way between 492~K and 518~K, hence we adopt $T_{\rm eff} = 505$~K, which implies log $g = 4.00$ (Table 5 and
Figure 4). A significantly warmer temperature is excluded: higher values of  $T_{\rm eff}$ result in too little flux at these wavelengths.
The photometry derived from a model with $T_{\rm eff} = 505$~K and log $g = 4.00$ is shown in Figures 6 and 7.
This result is consistent with the young disk kinematics, as the high value of $T_{\rm eff} = 550$~K corresponds to an age of 6.6~Gyr (Table 5), approaching the limit of what would be expected for the thin disk. However, the result does also mean that the models have an excess of flux at 1.0 -- 1.3~$\mu$m (Figure 5).  We estimate the uncertainty
in $T_{\rm eff}$ to be 10~K, by allowing for $\sim 2 \sigma$ variations in the W3 photometry and assuming a minimum mass of 2 -- 3~$M_{\rm Jupiter}$  for this brown dwarf. This uncertainty does not include systematic errors due to the known inadequacies of our model atmospheres (and all currently available model atmospheres) caused
by the incomplete opacities for atmospheres at these temperatures and pressures. 

In summary, we find that  UGPS 0722$-$05 has $T_{\rm eff} = 505 \pm 10$~K. This implies that the dwarf has a mass of 3 -- 11~$M_{\rm Jupiter}$ and an age between 60~Myr and 1 Gyr (see Table 5).  Figure 8 plots the entire spectral energy distribution for this solution, together with the observed spectra and observed and synthesized mid-infrared photometry.

\section{Alkali Chemistry at the T/Y Dwarf Boundary}

\subsection{Alkali Abundance}

In the atmospheres of late-type T dwarfs,
the refractory elements Mg, Al, Ca and Ti have been removed  by condensation 
at high temperatures in the deep atmosphere
and the Cs, K, Li, Na and Rb 
neutral alkalis are starting to be removed  as they 
gradually turn into chloride gases and other halide or hydroxide gases. 
Note that the fundamental vibrational frequencies of alkali halides are in the mid- to far-infrared, between 20 and 50~$\mu$m, where the low flux from the brown dwarf and the limited sensitivity of the instrumentation makes their detection extremely difficult.

Lodders (1999) gives a detailed description of the alkali chemistry in cool dwarf atmospheres, and here we summarize the major points that are important for understanding Rb and Cs chemistry. Rb and Cs lines, and the red wing of the strong 0.77~$\mu$m K I resonance doublet,
 are seen in our optical spectrum for the 500~K dwarf   UGPS 0722$-$05 in Figure 1 (see also Figures 10 and 11 below).

The trends in chemistry as function of pressure ($P$) and temperature ($T$) are shown in Figure 9, which also shows the $P$ - $T$ profiles for selected T dwarf atmosphere models, as indicated in the legends. Curves show where certain gases are equal in abundance --- for example Rb$ = $RbCl, long-dashed curve, indicates that the abundances of monatomic Rb equals that of RbCl gas. Other (solid) curves indicate where certain elements begin to be stable in condensed form --- for example $\rm Na_2S$(s) for sodium sulfide solid. 
The curves showing chemistry of Rb or Cs are colored red and blue, respectively.

The  Saumon \& Marley model atmospheres show that the Rb and Cs lines are formed in regions where the temperature is significantly warmer than the  $T_{\rm eff}$ value, and that the Rb I lines, by virtue of being closer to the K~I doublet, are formed on a higher background of opacity and so
higher up in the atmosphere in a cooler region than the Cs I lines. For a brown dwarf with  $T_{\rm eff} \approx 1000$~K, the Rb lines form at 930 -- 1000~K and the Cs lines at 1130 -- 1430~K. For 
$T_{\rm eff} \approx 750$~K these values become 880 -- 1000~K and 1050 -- 1230~K for Rb and Cs, respectively. For $T_{\rm eff} \approx 500$~K these values become 850 -- 940~K and 940 -- 1050~K for Rb and Cs, respectively. These zones are indicated along the  $P$ - $T$ profiles in Figure 9. For a  $T_{\rm eff} \approx 500$~K dwarf, like  UGPS 0722$-$05, we are just entering the temperature regime where the RbCl abundance is larger than the neutral Rb  abundance; the abundance of CsCl should be significantly larger than the neutral Cs  abundance.

For  $T_{\rm eff} \approx 500$~K brown dwarfs,  Figure 9 also indicates that $\rm Na_2S$ is condensing (the opacity of the solid is not currently included in the models). Removal of sodium from the gas reduces the abundances of monatomic Na and NaCl gas. Thus, there is more chlorine available to react with monatomic K and Rb gas to form KCl and RbCl gas and this accounts for the increase in KCl and RbCl only a few degrees below the onset of $\rm Na_2S$ condensation. 
With a further decrease in temperature, the partial pressures of the chloride gases increase and vapor saturation eventually sets in so that the K, Rb, and Cs chloride solid condensates become thermodynamically stable. 

Figure 10 shows our observed red spectra of  2MASS 0415$-$09 and  UGPS 0722$-$05; these dwarfs have $T_{\rm eff} \approx $  750~K and 500~K, and $\log g \approx$ 5.2 and 4.0, respectively (Saumon et al. 2007, this work). The  K I absorption at 0.770~$\mu$m remains significant for both T dwarfs, 
although the near-infrared 1.243/1.252~$\mu$m doublet becomes hard to detect for spectral types later than around T7 (e.g. McLean et al. 2003); the stronger 1.252~$\mu$m feature is marginally detected in 2MASS 0415$-$09   and  UGPS 0722$-$05 (McLean et al. 2003, Bochanski et al. 2011). The near-infrared features arise from a lower level  whose population  is very sensitive to temperature, and will disappear at warmer values  than the optical feature.
Neutral Cs and Rb are well detected in both objects, with similar line strengths in the two dwarfs, relative to the pseudo-continuum.  Thus although for Cs we are well into the regime where CsCl should be the dominant form, these lines are still seen in the far-red, where the atmosphere is relatively transparent. 

The persistence of Rb and Cs at lower temperature provides more evidence for  the existence of a deep silicate cloud layer (Ackerman \& Marley 2001; Burgasser et al. 2002b; Knapp et al. 2004; Marley et al. 2002; Marley, Saumon \& Goldblatt 2010). In strongly gravitationally bound atmospheres, condensates can settle into cloud layers below the cooler atmosphere, which then becomes depleted in the elements that are sequestered in the clouds.   
In brown dwarfs, removal of high temperature condensates into clouds prevents secondary condensate formation and solid solution formation. Thus,  the silicates  albite (Na-feldspar, NaAlSi$_3$O$_8$) and orthoclase (K-feldspar, KAlSi$_3$O$_8$) do not form, since the silicates which would 
form these compounds are sequestered in a cloud layer below 
the atmospheric level at which this reaction would otherwise proceed (Rb and Cs can also substitute in the feldspar lattice at temperatures lower than required for Na and K dissolution). Instead, Na condenses as sulfide, and the other alkalis condense as chlorides. 

Figure 11 compares the observed spectrum of UGPS 0722$-$05 to the spectrum generated by our $T_{\rm eff} = 505$~K model.  It can be seen that although the red wing of the  K I doublet at 0.770~$\mu$m is incorrectly modelled, the models reproduce the Rb feature reasonably well, while the modelled Cs lines are stronger than observed.
Now that brown dwarfs cooler than 500~K are being discovered (Cushing et al. 2011, Liu et al. 2011, Luhman et al. 2011), we can expect to see pronounced changes in the optical spectra of the coolest brown dwarfs, as  the alkalis convert into chloride gases and also condense as salts.

\subsection{Alkali Line Broadening and $iz$ Photometry}

Figures 12 and 13 show absolute $z$, $Y$, $J$ and $H$ magnitudes and colors, with model sequences.  The $iz$ data are AB magnitudes on the SDSS system (see \S  2.2.1), and the $YJH$ are Vega magnitudes on the Mauna Kea Observatories system (Tokunaga, Simons \& Vacca 2002, Tokunaga \& Vacca 2005). The photometry and parallaxes are taken from the literature, supplemented by the $iz$ photometry presented here. A compilation of $YJH$ data and parallaxes is available at \url{http://www.gemini.edu/staff/sleggett}, which is described in Leggett et al. (2010a) and references therein. Figures 12 and 13 include astrometry and photometry published after Leggett et al., in Marocco et al. (2010) and Burningham et al. (2010, 2011). Figures 12 and 13 also
include the published $z$ photometry described in \S 2.2.1, and SDSS $iz$ photometry for the brighter L and T dwarfs presented in  Geballe et al. (2002), Knapp et al. (2004) and Chiu et al. (2006).

Figure 12 implies that the calculated $i$ magnitude is too faint by $\sim$ 1 magnitude. This is consistent with
the discrepancy between the models and observations shown in Figure 11, where
the 0.770~$\mu$m  K I line is shallower than modelled ($i$ and $z$ bandpasses are shown in Figure 1). Figure 12 also suggests that $Y$ and $J$ are too bright by  $\sim$ 0.2 -- 0.3 magnitudes. The discrepancy at $Y$ and $J$ can be seen in Figure 5, where there is 
excess model flux at 1.0 -- 1.3~$\mu$m, which we interpret as due to the incompleteness of the CH$_4$ line list and possibly the presence of thin
clouds.

Marley et al. (2002, see also Burrows et al. 2000) point out that the $i - z$ color is sensitive to pressure broadening of the K I doublet. The exceptionally strong pressure broadening affecting the   0.770~$\mu$m K I resonance doublet requires the application of a sophisticated line
broadening theory. In the Saumon \& Marley models, these lines are modeled with a Lorentzian
core and a far wing exponential cutoff using a weakly constrained cutoff
parameter. The modeled $i - z$ color changes by as much as 0.4 magnitudes when
computed with the
cutoff parameter changed by a factor of two. A much more accurate
calculation of the
pressure-broadened line profiles of alkali metals in brown dwarfs has been
developed by Allard et al. (2003, 2005,
2007b) and Allard \& Spiegelman (2006) and will be incorporated in the  Saumon \& Marley
models.
The BT-Settl models (Allard et al. 2003, 2007a; Freytag et al. 2010), which
use these newer line profiles, do show a better match to the
UGPS 0722-05 optical spectrum.

Symbols in Figures 12 and 13 indicate the metallicity and gravity of the dwarfs, where they can be constrained by a companion, by bolometric luminosity, or by their IRAC colors (Leggett et al. 2011). There are no clear trends with metallicity or gravity, although the measurement uncertainties and scatter are large. The model sequences suggest that the $izYJ$ colors become more sensitive to  metallicity and gravity for $T_{\rm eff} \leq 700$~K, and trends may become clearer when more such objects, T9 and later-type dwarfs, are known.  $i - z$ and $z - J$ are calculated to be predominantly sensitive to metallicity, while $Y - J$ and $z - Y$ are sensitive to gravity. The colors of the latest-type T dwarfs, T9 -- T10, are consistent with models with  $T_{\rm eff} =$ 500 -- 600 ~K.

\section{Far-Red Colors and Spectral Type -- Brown Dwarfs in Surveys}

Figures 14 and 15 show $izYJ$ magnitudes and colors, respectively, as a function of spectral type, using the data sources described in \S 4.2. Also shown in Figure 14 are the detection limits, as apparent magnitudes, for LSST, Pan-Starrs and the VISTA Hemisphere Survey (VHS) (Ivezic et al. 2011, Magnier et al. 2009, \url{http://www.eso.org/sci/observing/policies/PublicSurveys/sciencePublicSurveys.html}). AB magnitudes have been converted to Vega magnitudes using the offsets given in Hewett et al. (2006).
The LSST and VHS surveys will cover 20,000 deg$^2$, Pan-Starrs will cover 30,000 deg$^2$. Due to the extremely red flux distribution of late-type T dwarfs and, presumably, early-type Y dwarfs, the $Y$ and $J$ filters give near-infrared surveys a significant advantage over the $i$- and $z$-bands of the far-red surveys. Together, the VHS and LSST surveys will enable cold dwarfs to be detected in both $Y$ and $J$, reducing the candidate contamination that occurs with single-band detections. The  final coadded depths of the LSST is expected to be  2.8 magnitudes deeper than indicated in Figure 14, so that VHS brown dwarfs will be detected in both LSST-$z$ and LSST-$y$ coadded images. The coadded LSST images will contain a larger number of brown dwarfs than the VHS, however these will predominantly be $y$-band only sources, requiring extensive followup. If the mass function is either log-normal or flat (Burningham et al. 2010, Covey et al. 2008, Kirkpatrick et al. 2011, Metchev et al. 2008, Reyle et al. 2010), then the simulations of Burgasser (2004) suggest that the VHS will enable the discovery of a few hundred 500~K brown dwarfs out to around 25~pc, and may find objects as cool as 350~K if they lie within 4~pc.

Figure 15 suggests that there is a large scatter in the far-red and near-infrared colors of brown dwarfs as a function of type. However the error bars are quite large, and the sample of brown dwarfs with $Y$-band data is biassed to types later than mid-T. The scatter seen in Figure 15 therefore may be a combination of measurement error and selection effects. 

\section{Conclusions}

We have supplemented the number of $iz$ measurements for brown dwarfs by obtaining photometry  for one L dwarf and 25 T dwarfs. We find that the $i - z$, $z - Y$ and $z - J$ colors of T dwarfs are very red, and continue to increase through to the late-type T dwarfs, with a hint of a saturation for
T8/10 types, or $T_{\rm eff} \approx 600$~K. Thus very cool brown dwarfs can be identified in far-red/near-infrared surveys such as  LSST, Pan-Starrs, UKIDSS and VISTA. Attention must be paid to photometric systems when working with the extreme energy distributions of cold brown dwarfs. Currently the Saumon \& Marley models do not reproduce the observed red wing of the strong 0.77~$\mu$m K I resonance doublet. The observed spectral shape is shallower, and the model calculated $i$ photometry is too faint by around 1 magnitude. New treatments of the line broadening will be incorporated in future work.

We have re-examined the flux distribution of the T10 dwarf UGPS 0722$-$05 (also classified as T9), using 
new optical and 3 -- 4 $\mu$m spectra, and new astrometry,  presented in this work.
We combine these results with the 
 near-infrared spectrum and photometry given by Lucas et al. (2010),  the mid-infrared photometry given by Lucas et al. and Kirkpatrick et al. (2011), and the radial velocity given by Bonchaski et al. (2011). We find that the bright 8 -- 16~$\mu$m flux  of UGPS 0722$-$05
excludes values of $T_{\rm eff}$ higher than around 505~K.  The luminosity we derive, using Saumon \& Marley models to determine bolometric corrections, is consistent with the far-red to mid-infrared flux distribution. We determine that UGPS 0722$-$05 has $T_{\rm eff} = 505 \pm 10$~K, a mass of 3 -- 11~$M_{\rm Jupiter}$ and an age between 60~Myr and 1 Gyr. This young age is consistent with the thin disk kinematics of the dwarf. 
The mass range overlaps with that usually considered to be
planetary, despite this being an unbound object discovered
in the field near the Sun; UGPS 0722$-$05 may be an example of the population of unbound Jupiter-mass objects identified in gravitational microlensing survey observations towards the Galactic Bulge (Sumi et al. 2011).
This apparently young rapid rotator is also undergoing vigorous vertical mixing
with a timescale of $\sim 1$~hr (described by a mixing coefficient of $K_{zz} = 10^{5.5}$ -- $10^{6}$ cm$^2$ s$^{-1}$), as determined by fits to the 4.5~$\mu$m IRAC and WISE photometry. The fit to the near-infrared spectrum is poor between 1.0 and 1.3 $\mu$m ---  a more complete line list for CH$_4$ opacity at these wavelengths is very much needed to understand the spectra of these ultracool dwarfs.

The new red spectrum of UGPS 0722$-$05 presented here shows that the neutral alkalis Cs, K and Rb continue to be readily detected even at  $T_{\rm eff}\approx 500$~K. This is because the red region of the spectrum is relatively clear, and the flux is emitted from a layer with brightness temperature 850 -- 1050~K.  The alkali chemistry of Lodders (1999) shows that 
this is a
temperature region where the abundances of the alkali chloride gases are equal to or
greater than those of the neutral alkali gases, and where $\rm Na_2S$ condenses. 
Hence for objects not much cooler than UGPS 0722$-$05 we should see the neutral alkalis weaken significantly, especially if temperatures become low enough for condensation of Rb and Cs chlorides.

\acknowledgments

SKL's research is  supported by Gemini Observatory.
The contribution of DS and MM was supported by NASA grant NNH11AQ54I. Work by KL was supported while working at the National Science Foundation.
Work by BF was supported by the National Science Foundation grant AST 0707377. 
ADJ is supported by a FONDECYT postdoctorado fellowship under project
number 3100098.

\clearpage

\begin{figure}
\includegraphics[angle=-90,scale=.6]{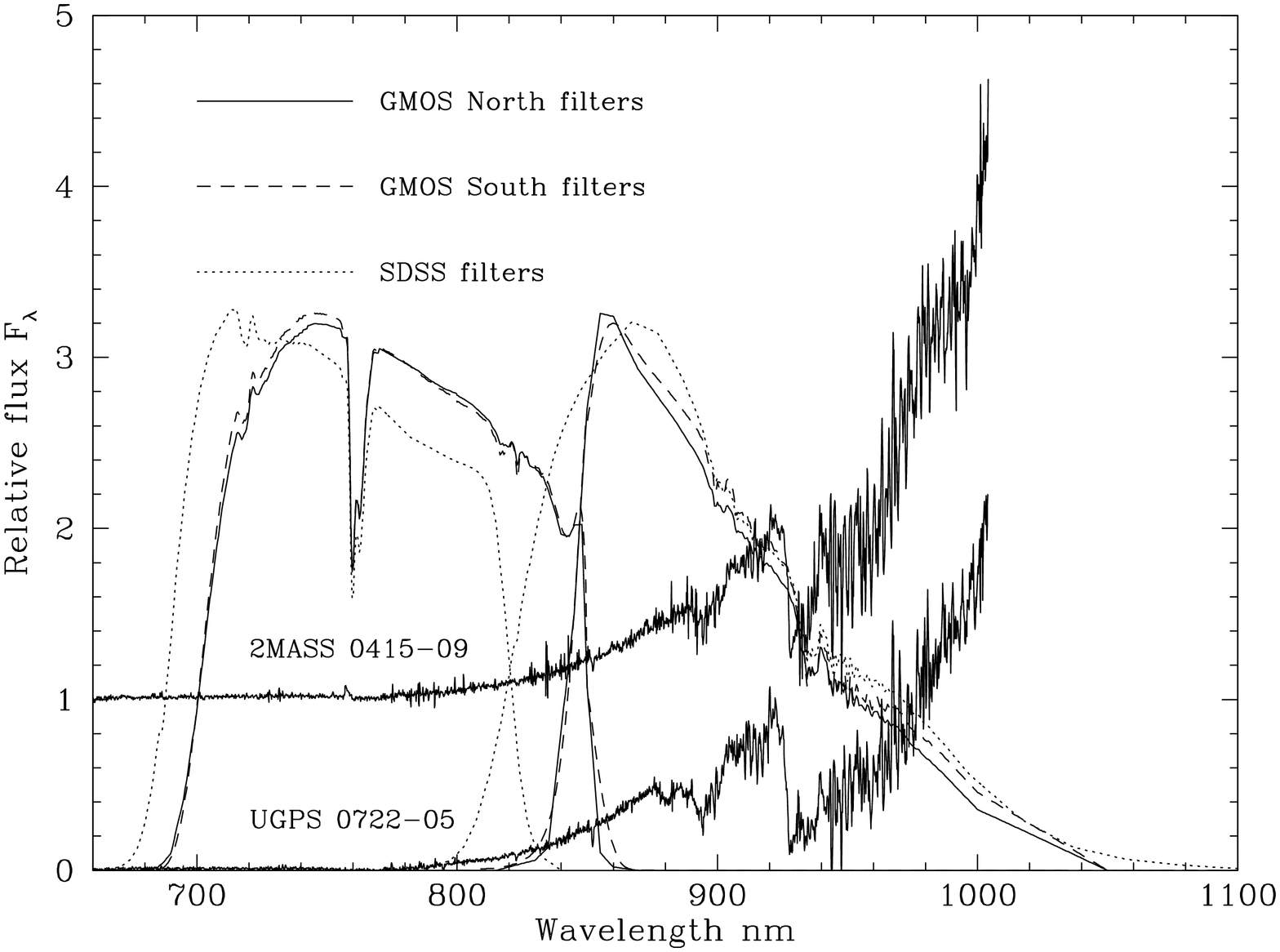}
\caption{GMOS spectra of 2MASS 0415$-$09 and UGPS 0722$-$05. The spectrum of 2MASS 0415$-$09 has been offset for clarity, and both spectra have been normalized to their flux at 920~nm. The GMOS North and South $iz$ filter profiles are shown, as well as the SDSS profiles. These include filter and optics transmission, detector response, and atmospheric transmission. The SDSS values have been taken from Doi et al. 2010; the GMOS values can be found on the Gemini web pages: 
http://www.gemini.edu/sciops/instruments/gmos/imaging?q=node/10415,
http://www.gemini.edu/sciops/instruments/gmos/imaging/filters,
http://www.gemini.edu/sciops/instruments/gmos/imaging/detector-array.
The filter transmission values have been arbitrarily scaled for plotting purposes. 
\label{fig1}}
\end{figure}

\clearpage

\begin{figure}
\includegraphics[angle=-90,scale=.6]{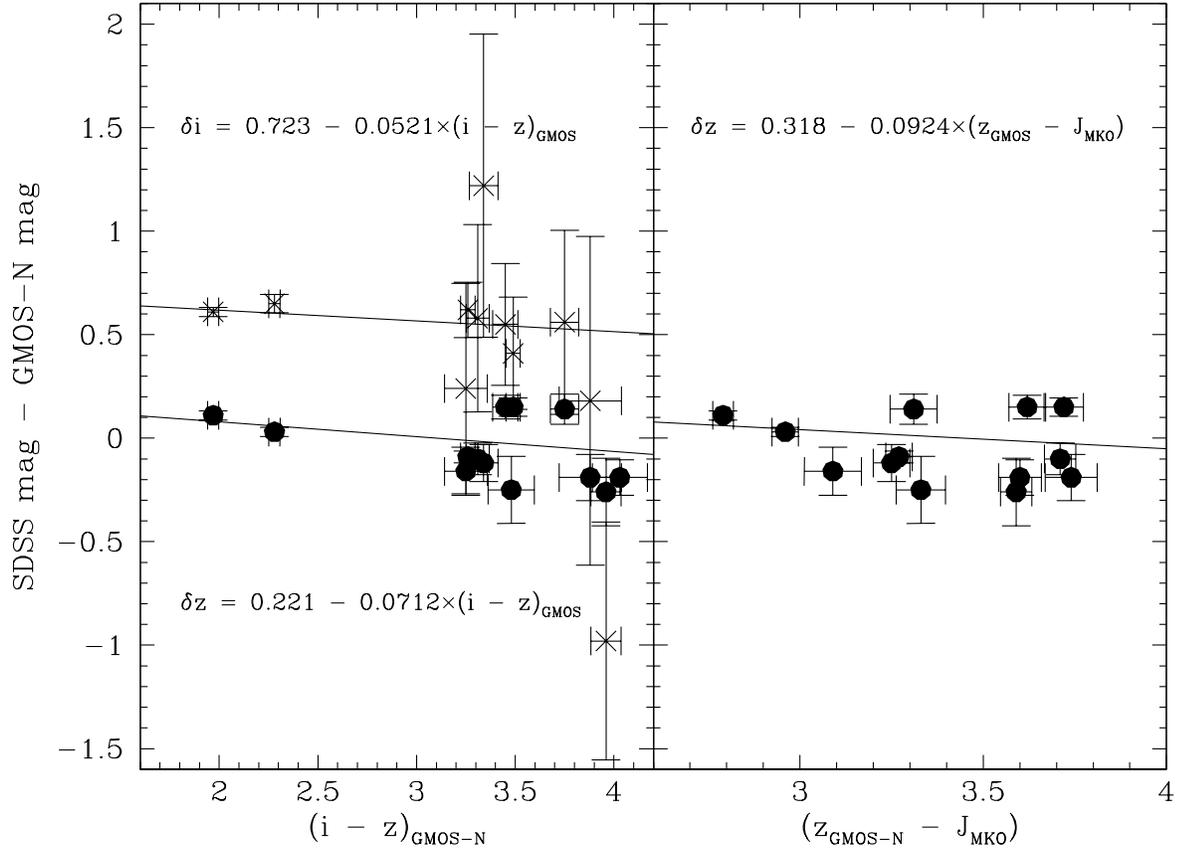}
\caption{Color transformations between the GMOS and SDSS systems. Filled circles are $\delta z$ (SDSS-z - GMOS-N-z) and crosses $\delta i$ (SDSS-i - GMOS-N-i). The large uncertainties in some of the $\delta i$ are due to the large uncertainties in the measured SDSS $i$ magnitudes for these relatively faint sources (see Table 3).
\label{fig2}}
\end{figure}

\begin{figure}
\includegraphics[angle=-90,scale=.7]{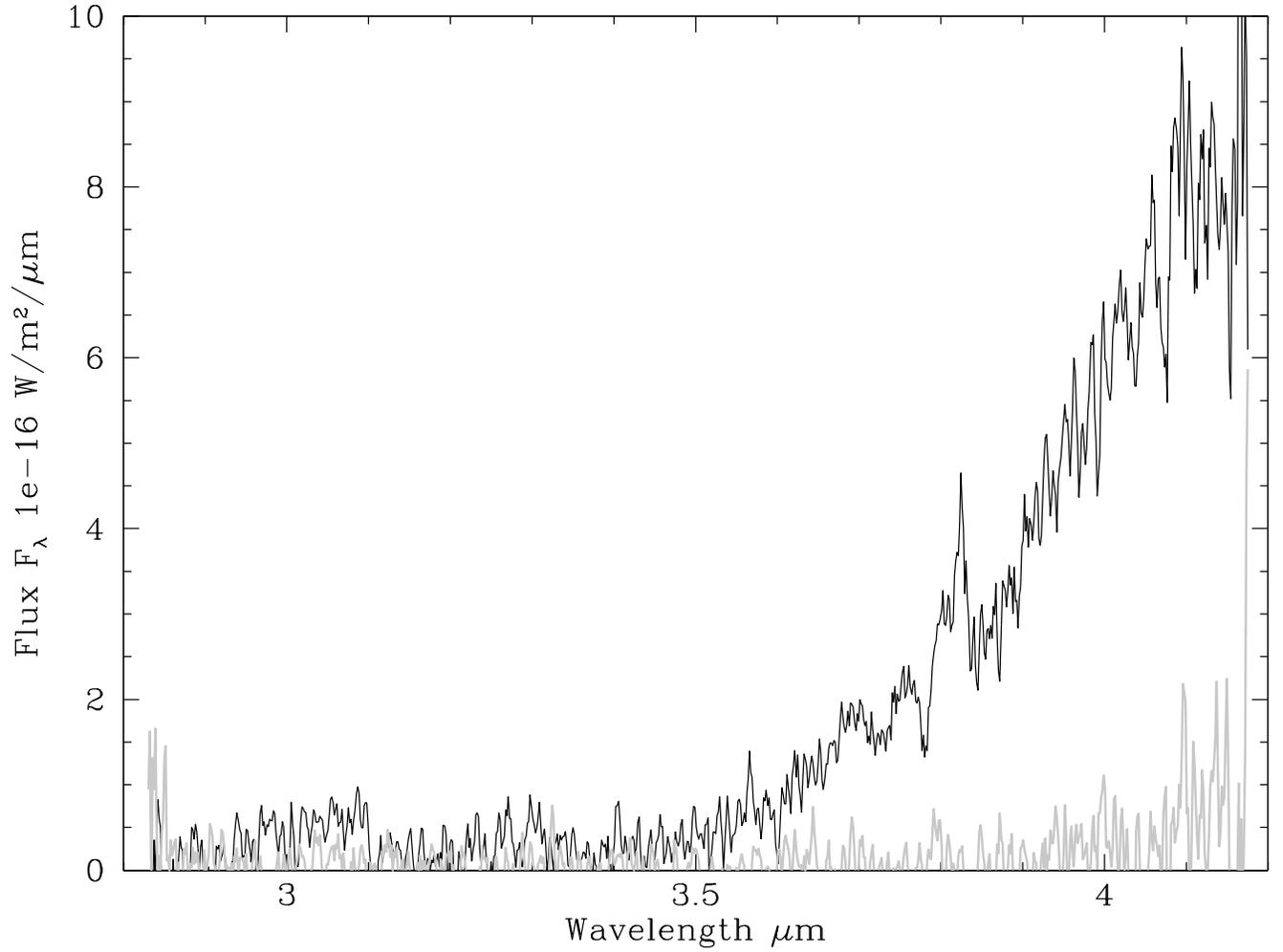}
\caption{UGPS 0722$-$05 2.8 -- 4.2~$\mu$m spectrum obtained using the Infrared Camera and
Spectrograph on Subaru, median-smoothed by 3 pixels. The error spectrum is shown below, in grey. Absorption by CH$_4$ dominates the spectrum (see Figure 5).
\label{fig3}}
\end{figure}

\clearpage
\begin{figure}
\includegraphics[angle=0,scale=.5]{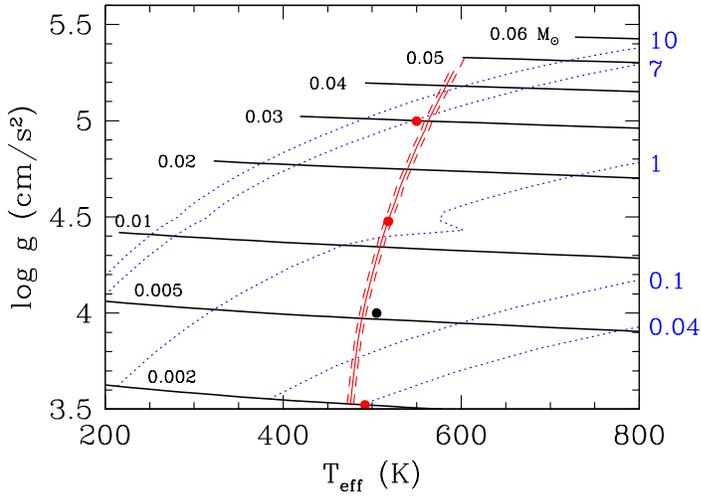}
\caption{Solutions obtained with our method whereby the luminosity implied by the
spectrum agrees with that implied by the evolution for UGPS 0722$-$05. The solutions 
lie on a line of approximately constant luminosity in the ($T_{\rm eff}$, log $g$) plane, along which we have picked 3 representative points
shown by the red dots (see Table 5).  The black dot is the solution that
corresponds to our best compromise fit of all the spectroscopic and
photometric
data (see text). The solid and dashed line show a constant value of
log $L_{bol}/L_{\odot} = -6.13 \pm 0.014$, corresponding to the $T_{\rm eff}=518\,$K
solution.   Cooling tracks for the cloudless evolution
of Saumon \& Marley (2008), labeled by the mass in solar units, are shown in
black. Isochrones (dotted blue lines) are labeled in Gyr on the right side
of the graph.
\label{fig4}}
\end{figure}

\clearpage
\begin{figure}
\includegraphics[angle=0,scale=.6]{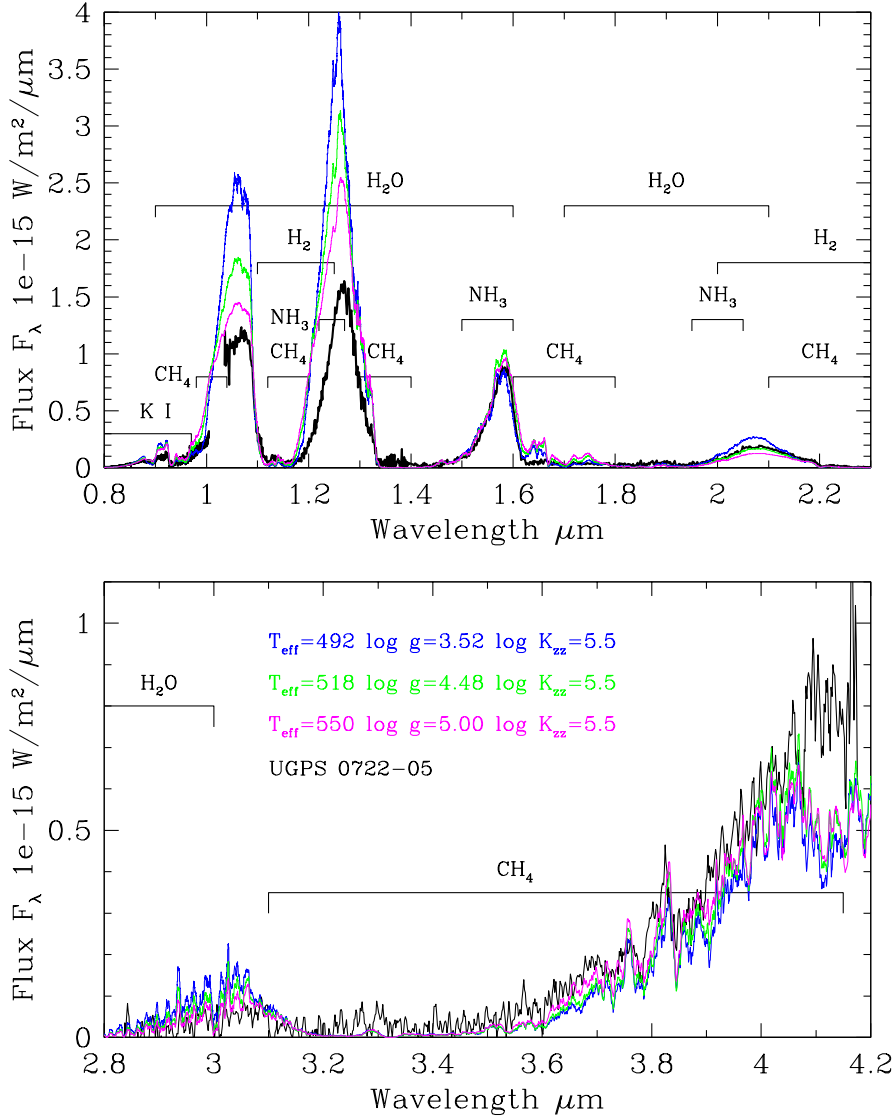}
\caption{Comparison of observed (this work and the near-infrared spectrum of Lucas et al. 2010) and modelled spectra for  UGPS 0722$-$05.  Principal opacity sources are indicated; the location of NH$_3$ is taken from the line identifications of Bochanski et al. (2011). These Saumon \& Marley models are cloudless, solar-metallicity models which include non-equilibrium chemistry; the legend shows the $T_{\rm eff}$, $\log g$ and  $\log K_{zz}$ values for each model. The synthetic spectra are calibrated using the measured distance to the object, and the radius implied by evolutionary models for each ($T_{\rm eff}$, $\log g$). The discrepancy at 1.0 -- 1.3~$\mu$m, and 4.1~$\mu$m,  is likely caused by incomplete
 CH$_4$ opacity line lists although thin clouds may also affect the relative
peak
fluxes in the near-infrared. 
\label{fig5}}
\end{figure}

\clearpage
\begin{figure}
\includegraphics[angle=-90,scale=.6]{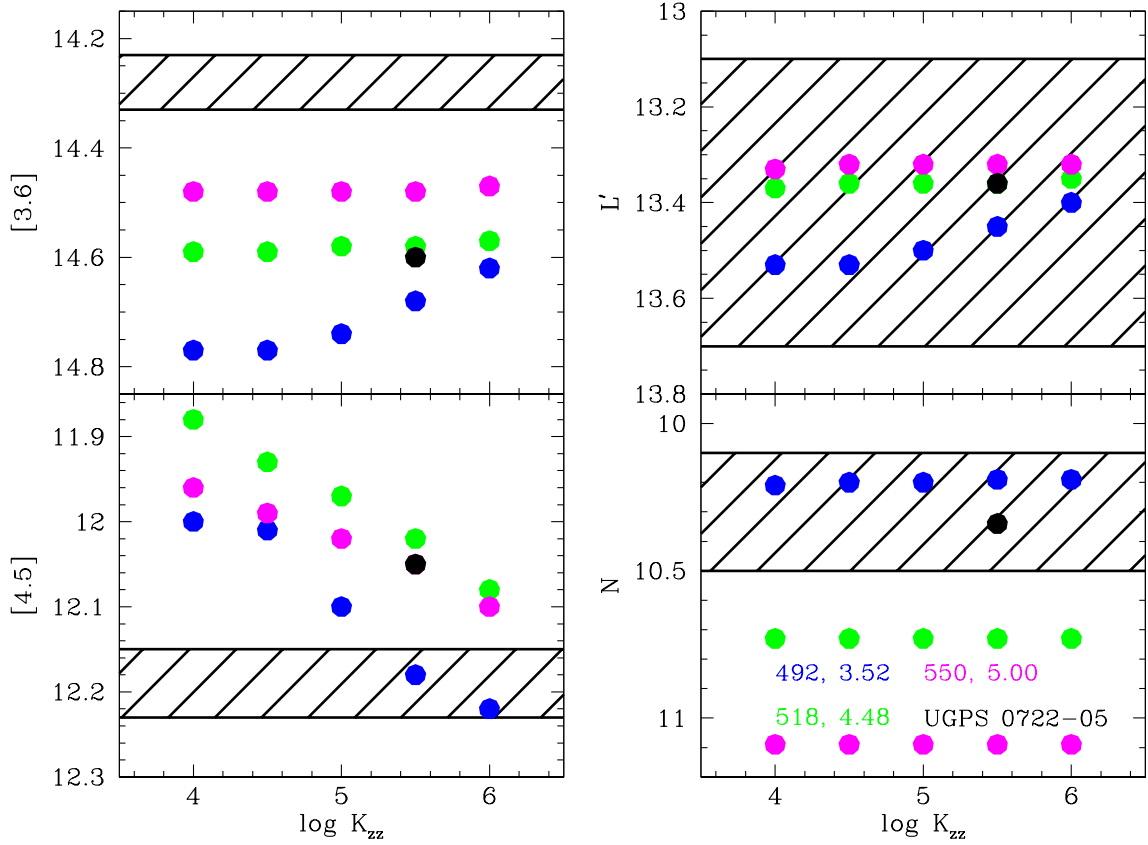}
\caption{Comparison of observed and modelled apparent magnitudes as a function of the vertical mixing coefficient $K_{zz}$ cm$^2$ s$^{-1}$. The $1 \sigma$ range in the measured IRAC [3.6] and [4.5], $L^{\prime}$ and $N$ magnitudes (Lucas et al. 2010) is indicated by the shaded rectangles in each panel (the uncertainty in the $L^{\prime}$ value is large). Synthetic photometry from three Saumon \& Marley models are shown as colored dots; the legend indicates the ($T_{\rm eff}$, $\log g$) values for each model (see Table 5). The black dot indicates the value for our adopted $T_{\rm eff} = 505$~K, log $g = 4.00$, log $K_{zz} = 5.5$, solution.
\label{fig6}}
\end{figure}

\clearpage
\begin{figure}
\includegraphics[angle=-90,scale=.6]{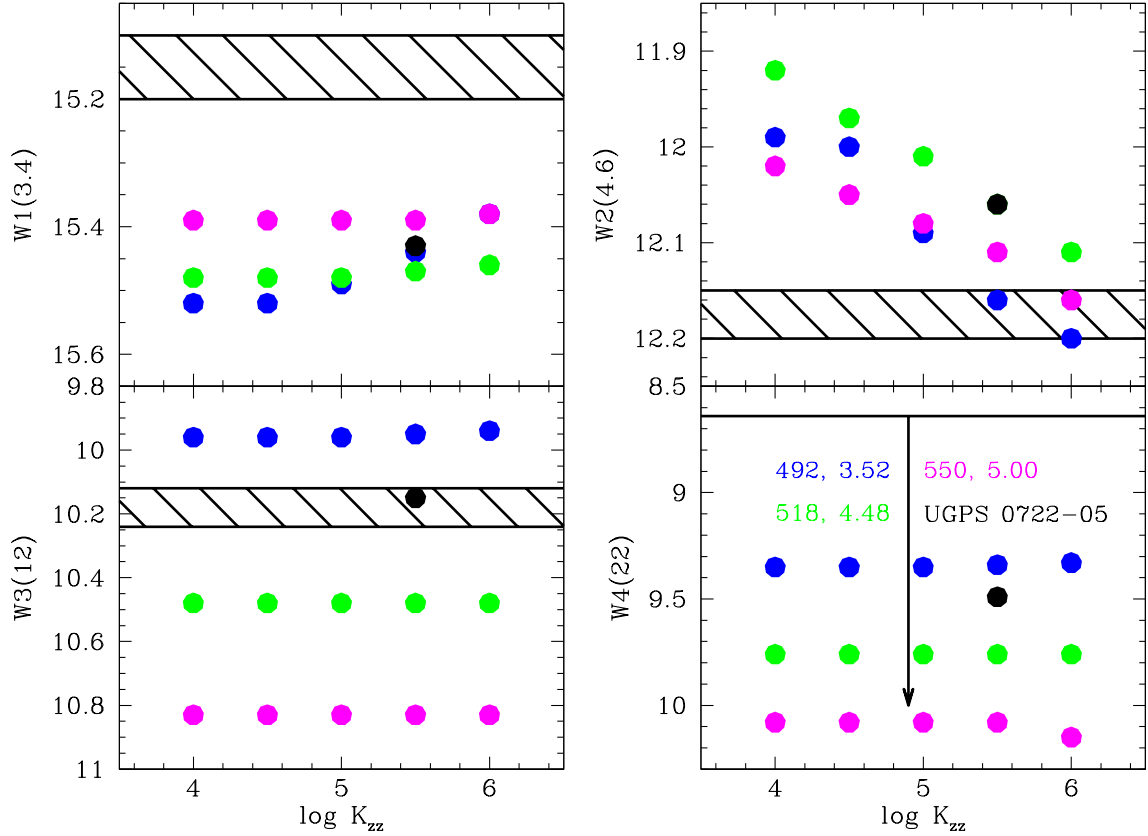}
\caption{Comparison of observed and modelled apparent magnitudes as a function of the vertical mixing coefficient $K_{zz}$ cm$^2$ s$^{-1}$. The $1 \sigma$ range in the measured WISE  W1(3.4), W2(4.6) and W3(12)  magnitudes (Kirkpatrick et al. 2011)   is indicated by the shaded rectangles in each panel, an upper limit is indicated for W4(22). Symbols are as in Figure 6.
\label{fig7}}
\end{figure}

\clearpage
\begin{figure}
\includegraphics[angle=-90,scale=.6]{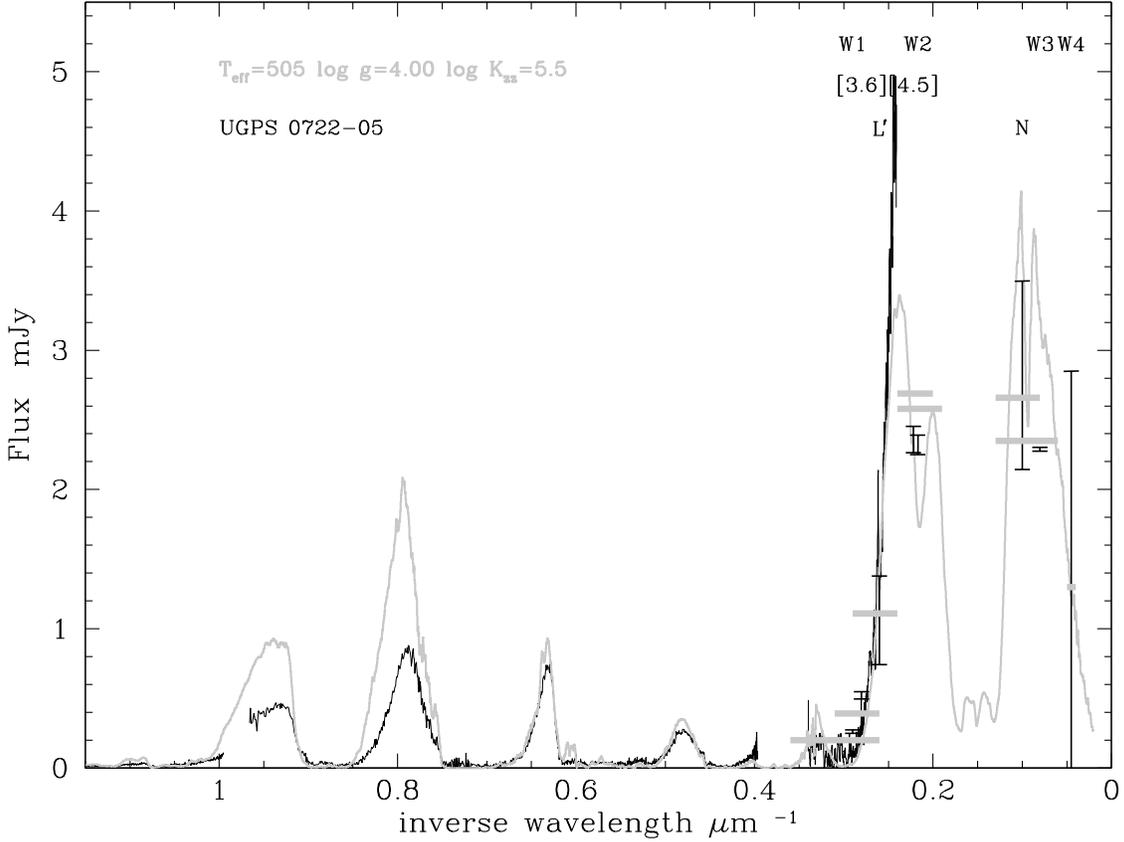}
\caption{Comparison of the observed and modelled complete spectral energy distribution for  UGPS 0722$-$05. Black curves are the observed far-red and 3 -- 4~$\mu$m spectra presented in this work, and the near-infrared spectrum of Lucas et al. (2010). Black error bars represent the observed 3 -- 16~$\mu$m photometry, with an upper limit indicated for the 22~$\mu$m WISE datapoint. The grey curve is our adopted model, as described in the text, scaled by the measured distance and the radius implied by evolutionary models. The grey horizontal lines indicate the photometry computed from this model over the IRAC, WISE and ground-based $L^{\prime}$ and $N$ filter bandpasses. 
\label{fig8}}
\end{figure}

\clearpage
\begin{figure}
\includegraphics[angle=0,scale=.6]{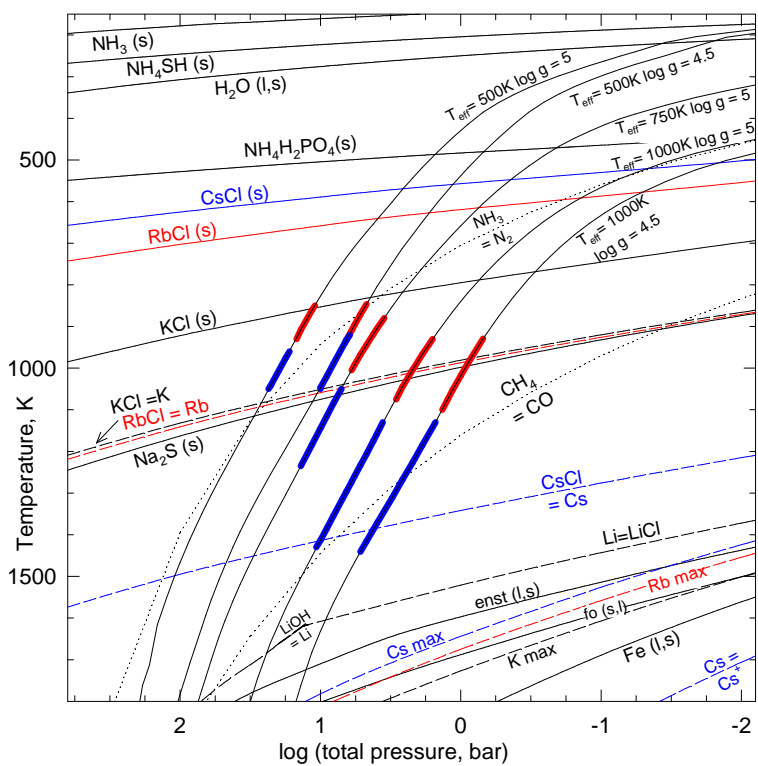}
\caption{Pressure($P$)-Temperature($T$) profiles for T dwarf atmospheres with 
 $T_{\rm eff}$ and log $g$ as indicated in the legends.
Curves show where certain gases are equal in abundance --- for example Rb = RbCl, long-dashed curve, implies that the abundances of monatomic Rb equals that of RbCl gas. 
Other (solid) curves indicate where certain elements begin to be stable in condensed form -- for example $\rm Na_2S$(s) for sodium sulfide solid. The red and blue zones along the $P$ - $T$ profiles  indicate the regions where the Rb and Cs lines form, respectively. Note these regions are hotter than the $T_{\rm eff}$ value.
\label{fig9}}
\end{figure}

\clearpage
\begin{figure}
\includegraphics[angle=-90,scale=.6]{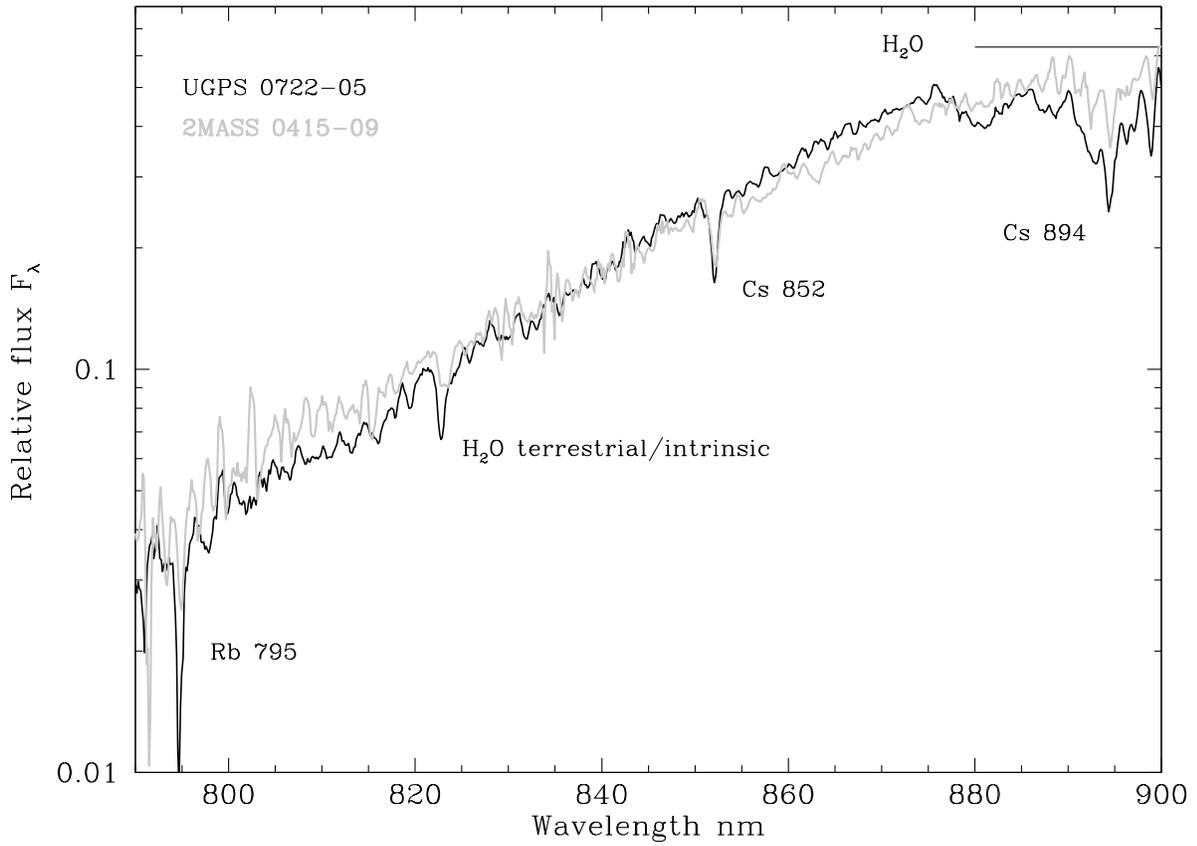}
\caption{GMOS spectra of 2MASS 0415$-$09 (grey curve) and UGPS 0722$-$05 (black curve), normalized to their flux at 920~nm, and boxcar smoothed. 
\label{fig10}}
\end{figure}

\clearpage
\begin{figure}
\includegraphics[angle=-90,scale=.6]{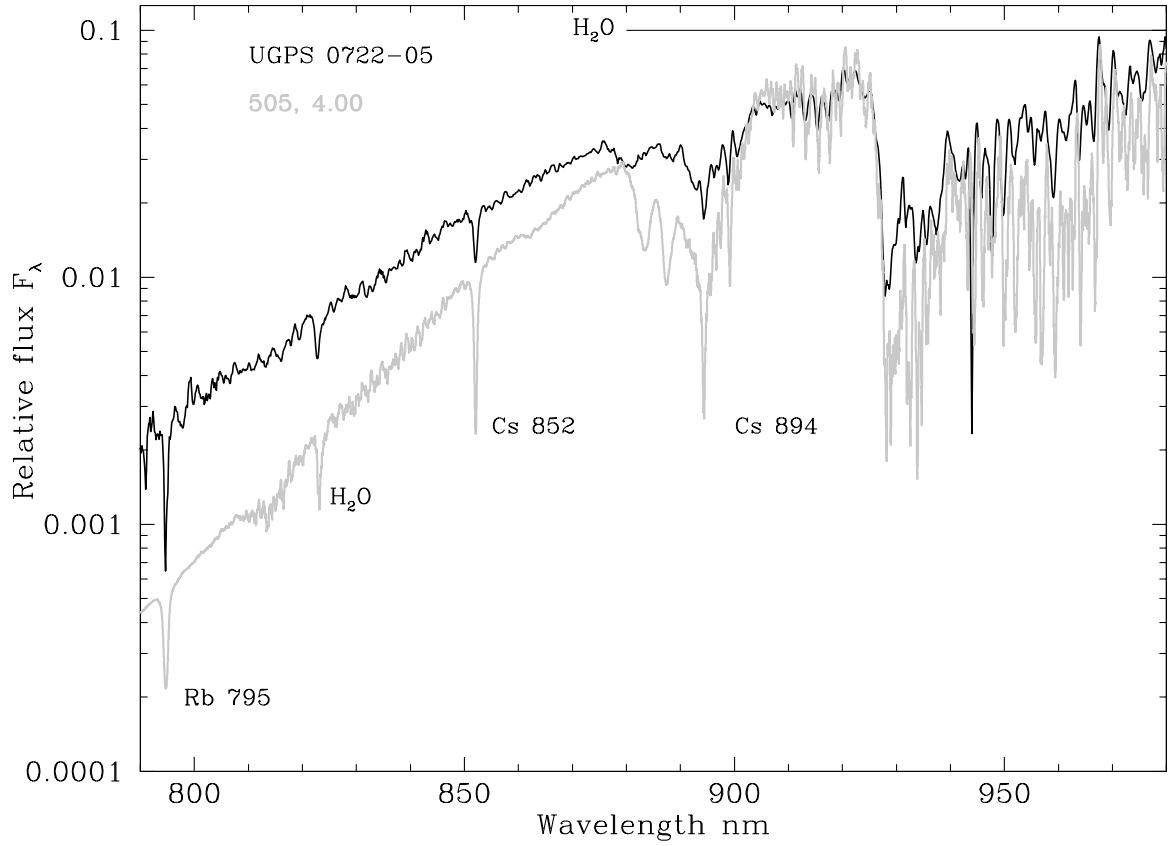}
\caption{Observed (black curve) and modelled (grey curve) spectra of UGPS 0722$-$05, normalized to their flux at 920~nm. The red wing of the 0.77~$\mu$m K~I doublet is incorrectly modeled, see the discussion in \S 4.2 of the text.  
\label{fig11}}
\end{figure}

\clearpage
\begin{figure}
\includegraphics[angle=-90,scale=.6]{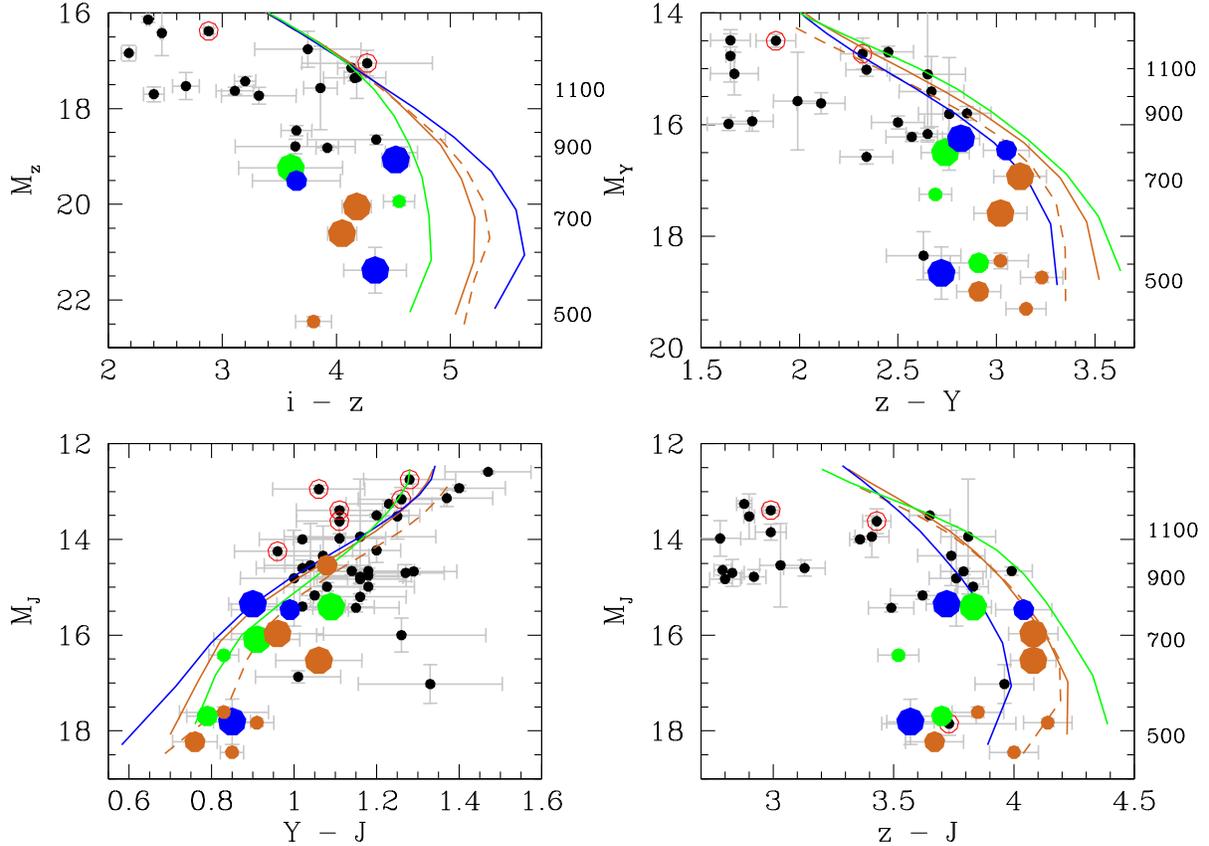}
\caption{Absolute magnitudes as a function of color. The $iz$ data are AB magnitudes on the SDSS system (\S  2.2.1), and the $YJH$ are Vega magnitudes on the Mauna Kea Observatories system (Tokunaga, Simons \& Vacca 2002, Tokunaga \& Vacca 2005). Known binaries are circled in red, the photometry is for the unresolved system. Larger colored symbols indicate metallicity and gravity, where known. Large, medium and small colored circles represent $\log \, g \approx$ 5.0--5.5, 4.5--5.0 and  4--4.5. Green circles are metal-rich, orange have solar metallicity, and blue are metal-poor. Sequences from the Saumon \& Marley models are also shown. Solid lines are $\log \, g = 4.48$, dashed are $\log \, g = 5.0$. Green lines are [m/H]$=+0.3$, blue [m/H]$=-0.3$, and orange  [m/H]$=0$. $T_{\rm eff}$ values for the [m/H]$=0$  $\log \, g = 4.48$ model are shown on the right axis.  For $T_{\rm eff}$ 500 -- 1100~K,
$\log \, g = 4.48$ corresponds to 10 -- 15 $M_{\rm Jupiter}$ and ages 2 -- 0.2 Gyr, respectively (see Figure 4).
\label{fig12}}
\end{figure}

\clearpage
\begin{figure}
\includegraphics[angle=-90,scale=.6]{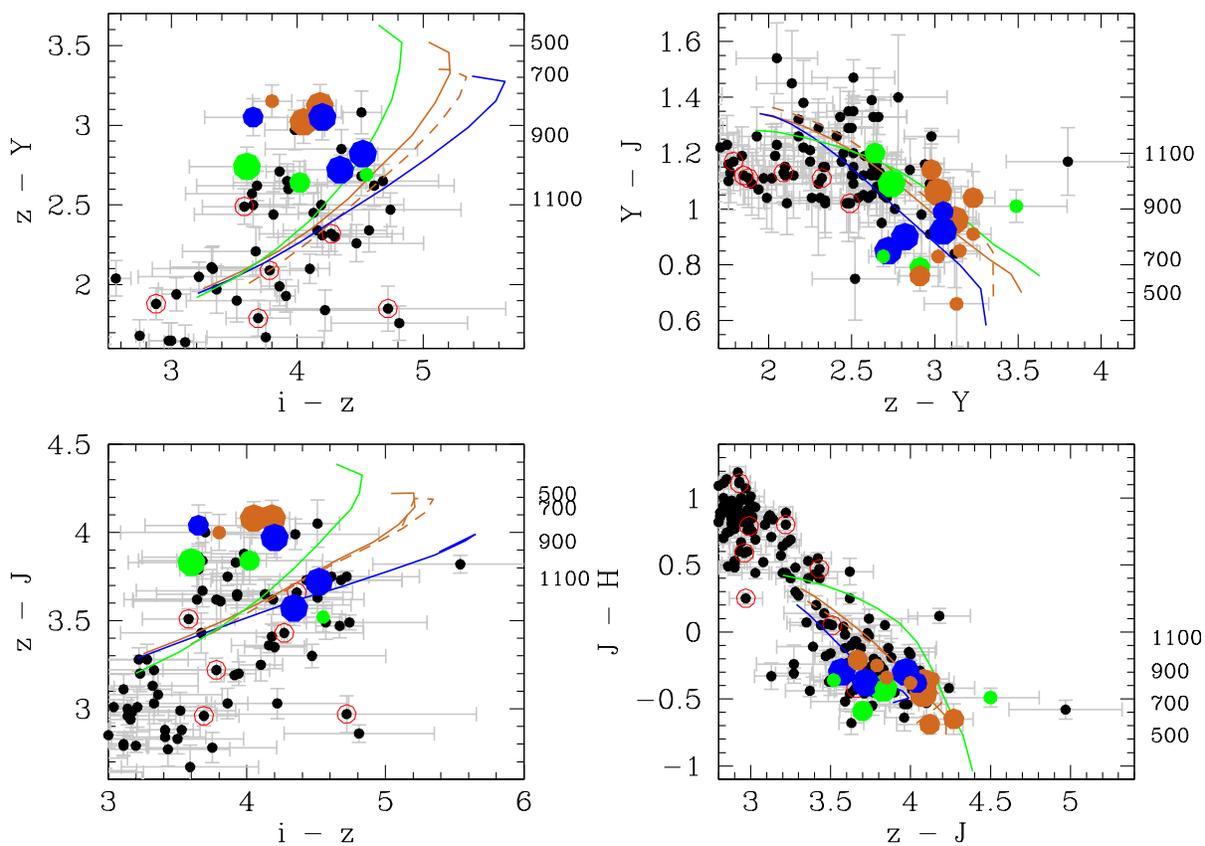}
\caption{Far-red and near-infrared color-color plots. Symbols and sequences are as in Figure 12. The $iz$ data are AB magnitudes on the SDSS system (\S  2.2.1), and the $YJH$ are Vega magnitudes on the Mauna Kea Observatories system (Tokunaga, Simons \& Vacca 2002, Tokunaga \& Vacca 2005).
\label{fig13}}
\end{figure}
\clearpage

\clearpage
\begin{figure}
\includegraphics[angle=-90,scale=.6]{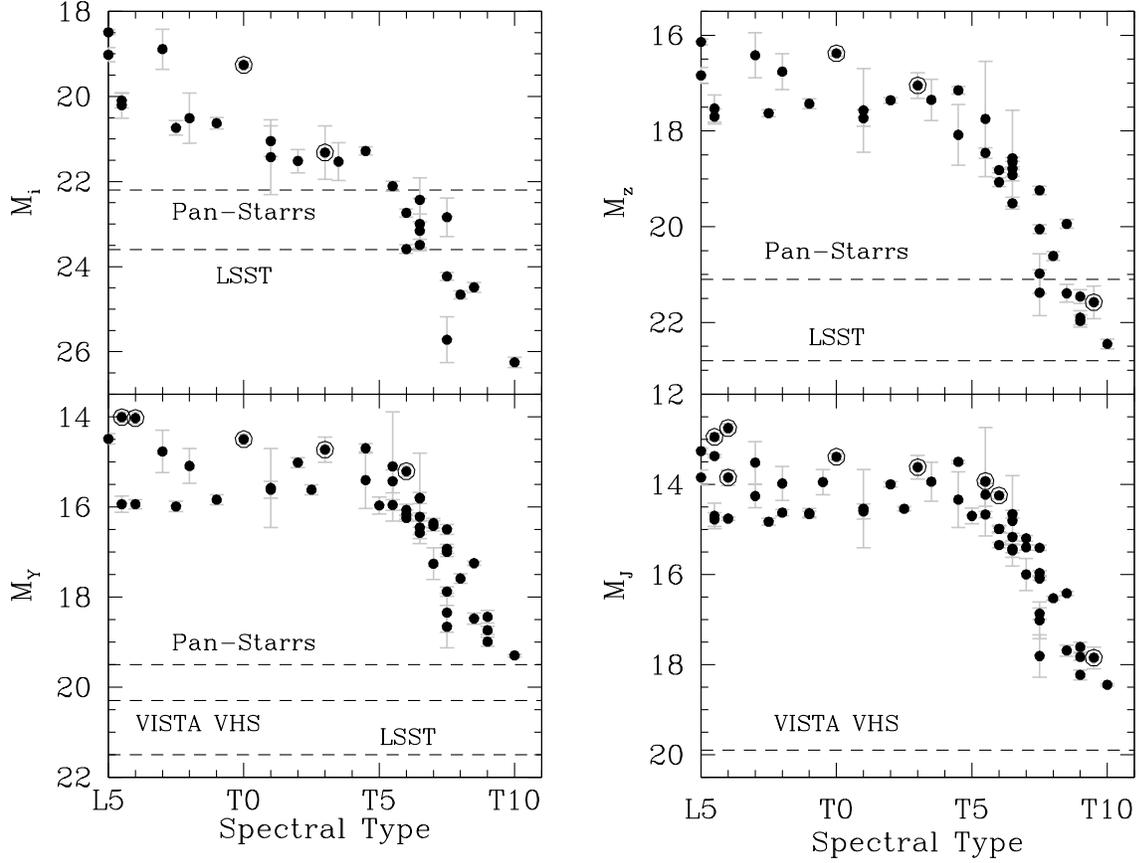}
\caption{Absolute magnitudes as a function of spectral type. The $iz$ data are AB magnitudes on the SDSS system (\S  2.2.1), and the $YJH$ are Vega magnitudes on the Mauna Kea Observatories system (Tokunaga, Simons \& Vacca 2002, Tokunaga \& Vacca 2005). Known binaries are circled, the photometry is for the unresolved system. Detection limits for the LSST (single exposure), Pan-Starrs and VISTA-VHS surveys are indicated, as apparent magnitudes. These limits indicate what spectral types can be detected at a distance of 10~pc. 
\label{fig14}}
\end{figure}

\clearpage
\begin{figure}
\includegraphics[angle=-90,scale=.6]{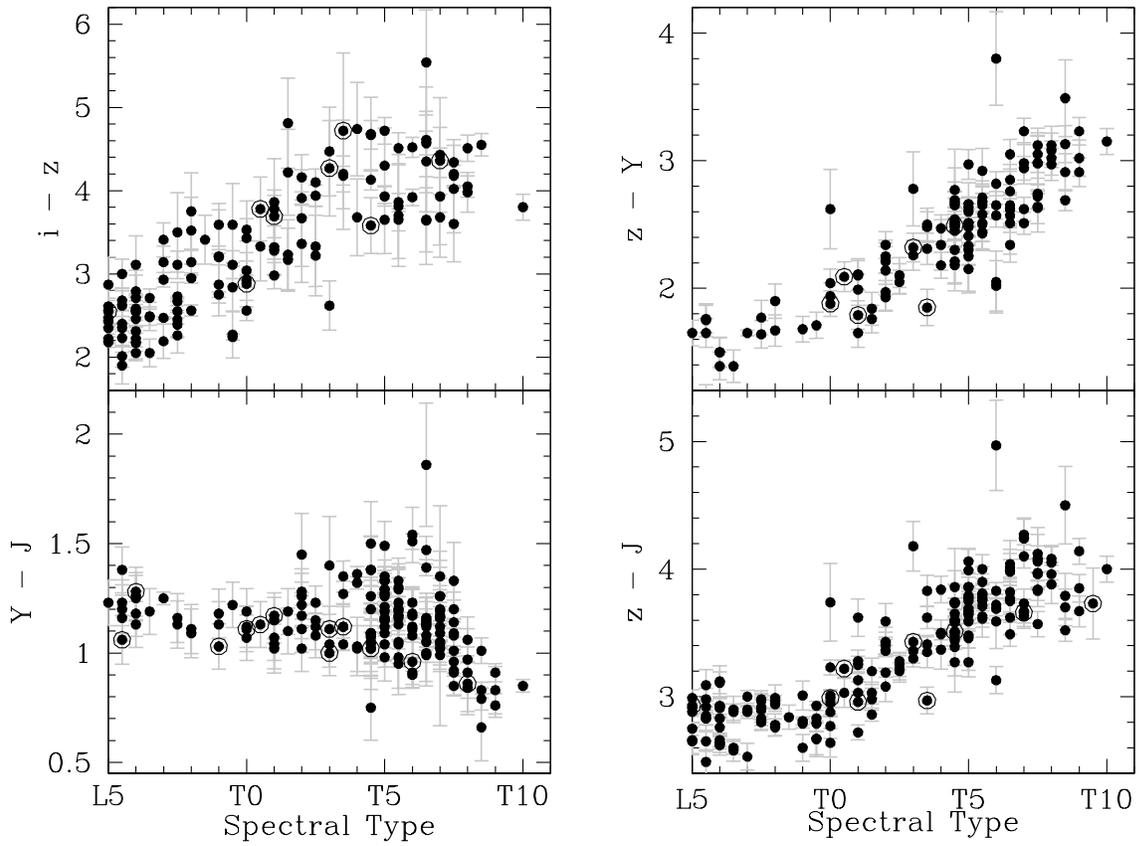}
\caption{Colors as a function of spectral type. Known binaries are circled, the photometry is for the unresolved system. The $iz$ data are AB magnitudes on the SDSS system (\S  2.2.1), and the $YJH$ are Vega magnitudes on the Mauna Kea Observatories system (Tokunaga, Simons \& Vacca 2002, Tokunaga \& Vacca 2005).
\label{fig15}}
\end{figure}
\clearpage

\begin{deluxetable}{lrccl}
\tabletypesize{\footnotesize}
\tablecaption{GMOS Spectroscopic Observation Log}
\tablewidth{0pt}
\tablehead{
\colhead{Short Name} & \colhead{SDSS $z$} & \colhead{Total Exp.} & \colhead{Date} & Program \\
\colhead{}  & \colhead{AB}  &\colhead{hours} & \colhead{YYYYMMDD} & \\
}
\startdata
2MASS 0415$-$09 & 19.40$\pm$0.06 & 1.5 & 20100905, 20100906 & GN-2010B-Q-59 \\
UGPS 0722$-$05 & 20.52$\pm$0.10 & 6.0 & 20101103, 20101107, 20101108 & GN-2010B-Q-59 \\
\enddata
\end{deluxetable}

\clearpage

\begin{deluxetable}{lcrrcclc}
\tabletypesize{\scriptsize}
\tablewidth{0pt}
\rotate
\tablecaption{GMOS $i$ and $z$ Photometry for L and T  Dwarfs on Each GMOS Natural System}
\tablehead{
\colhead{Name} & \colhead{Spectral} & \colhead{$i$(err)} & \colhead{$z$(err)} & \colhead{Exp. $i$, $z$}   & \colhead{Date} & \colhead{Program} & \colhead{Discovery}\\
\colhead{RA/Dec.} & \colhead{Type} &\colhead{} & \colhead{}  & \colhead{minutes} & \colhead{YYYYMMDD} & \colhead{} & \colhead{Reference{\tablenotemark{a}}}\\
}
\startdata
2MASSI J0415195$-$093506   & T8 & 22.93(0.04)  & 19.46(0.03)  & 20, 2 & 20100103  & GN-2010A-Q-81 &  1\\
2MASSI J0415195$-$093506   & T8 & 22.93(0.05) &  19.55(0.06) &  16, 1.2 & 20101115  & GS-2010B-Q-39  & 1\\
SDSSp J042348.57$-$041403.5  & T0 & 19.54(0.02) &  17.26(0.02) & 2, 1.3 & 20100313, 20100315  & GN-2010A-Q-81& 2\\
SDSSp J042348.57$-$041403.5  &  T0 &  19.58(0.02) &  17.32(0.02) &  0.7, 0.3 & 20101106  & GS-2010B-Q-39 & 2\\
2MASS J05591914$-$1404488  & T4.5 & 20.77(0.05) &  17.27(0.05)  & 6.7, 0.7 & 20100402  & GN-2010A-Q-81 &  3\\
2MASS J05591914$-$1404488  & T4.5 & 20.87(0.03)  & 17.28(0.02)  & 1.3, 0.3 & 20101103  & GS-2010B-Q-39 & 3\\
UGPS J072227.51$-$054031.2  & T10 & 23.76(0.09)  & 20.58(0.06)  & 80, 2 & 20100315  &   GN-2010A-Q-81&  4\\
2MASS J07290002$-$3954043   & T8 &  23.71(0.09)  & 19.78(0.03) & 16, 0.7  & 20110109 & GS-2010B-Q-39 & 5\\
SDSS J075840.33$+$324723.4  & T2 & 21.31(0.03) &  18.05(0.02)  & 2, 1.3 & 20100315 & GN-2010A-Q-81&  6\\
SDSS J083048.80$+$012831.1  & T4.5 & 23.67(0.13)  & 19.64(0.05)  &  32, 1.3 & 20110126,20110214 & GS-2010B-Q-39&  6\\
SDSSp J083717.22$-$000018.3  & T1 & 23.29(0.09)  & 20.04(0.06)  & 8, 0.3 & 20110110 & GS-2010B-Q-39 & 7\\
ULAS J092624.76$+$071140.7   & T3.5 &   24.55(0.16) &  20.87(0.08)  & 90, 3 & 20110110 & GS-2010B-Q-39 &  8\\
2MASSI J0937347$+$293142  &  T6 & 22.01(0.03)  & 18.04(0.02)  & 6.7, 0.7 & 20100506 & GN-2010A-Q-81&  1\\
ULAS J095047.28$+$011734.30  & T8 & 25.41(0.19)  & 22.03(0.07)  & 112, 13 & 20110111,20110114 & GS-2010B-Q-39 & 9\\
2MASSI J1047538$+$212423 & T6.5 & 23.06(0.07)  & 19.00(0.03)  & 16, 0.7 & 20101214 & GS-2010B-Q-39 & 10\\
SDSS J104829.21$+$091937.8   &  T2.5 & 23.03(0.06)  & 19.69(0.04)  & 16, 0.7 & 20110110 & GS-2010B-Q-39 & 11\\
SDSSp J111010.01$+$011613.1  & T5.5 &  23.79(0.15) &  19.91(0.05)  & 54, 3 & 20110126 & GS-2010B-Q-39 & 2\\
2MASS J11145133$-$2618235 & T7.5  &   23.21(0.09) &  19.59(0.04)  & 16, 0.7 & 20110125 & GS-2010B-Q-39 & 12\\
ULAS  J115759.04$+$092200.7  & T2.5 & 23.70(0.10) &  20.22(0.06)  & 32, 1.3 & 20110109 & GS-2010B-Q-39 & 13\\
2MASS J12314753$+$0847331  & T5.5 & 22.26(0.05) &  18.81(0.04)  & 16, 0.7 & 20110126 & GS-2010B-Q-39 & 14\\
ULAS J130041.73$+$122114.7  &  T8.5 & 24.26(0.07)  & 20.30(0.03) & 67.5, 4.5 &  20100313 & GN-2010A-Q-81 & 15\\
SDSS J141624.08$+$134826.7A{\tablenotemark{b}} & L7  & 17.77(0.02)  & 15.80(0.02) & 72, 2 &  20100705, 20100712 & GN-2010A-Q-81 & 16\\
SDSS J141623.94$+$134836.3B{\tablenotemark{b}} & T7.5 & 24.69(0.25)  & 20.88(0.05) & 72, 2 &  20100705, 20100712 & GN-2010A-Q-81 &  17\\
Gliese 570D (2MASS J14571496$-$2121477)  & T7.5 & 22.54(0.05) &  18.96(0.03)  & 6.7, 0.7 & 20100315 & GN-2010A-Q-81&  18\\
SDSSp J162414.37$+$002915.6  & T6 & 22.41(0.03)  & 18.92(0.02) & 20, 2 &  20100412 & GN-2010A-Q-81 & 19\\
SDSSp J175032.96$+$175903.0  &  T3.5 & 23.20(0.06)  & 19.45(0.04)  & 20, 2 & 20100411, 20100413 & GN-2010A-Q-81 & 2\\
Wolf 940B (ULAS J214638.83$-$001038.7){\tablenotemark{c}}  & T8.5 &  &  21.91(0.11)  & -, 9 & 20100621, 20100703 & GN-2010A-Q-81 & 20\\
2MASSI J2339101$+$135230  & T5 & 22.88(0.05) &  19.57(0.03) & 16, 0.7 &  20101103 & GS-2010B-Q-39 &  1\\
2MASSI J2356547$-$155310  & T5.5 & 22.41(0.04) &  19.36(0.03)  & 8, 0.3 & 20101103 & GS-2010B-Q-39 & 1\\
\enddata
\tablenotetext{a}{Discovery references are: 1 - Burgasser et al. 2002a; 2 - Geballe et al. 2002; 3 - Burgasser et al. 2000b;  4 - Lucas et al. 2010; 5 - Looper et al. 2007; 6 - Knapp et al. 2004; 7 - Leggett et al. 2000; 8 - Burningham et al. 2010; 9 - Burningham et al. in prep.; 10 - Burgasser et al. 1999; 11 - Chiu et al. 2006; 12 - Tinney et al. 2005; 13 - Pinfield et al. 2008; 14 - Burgasser et al. 2004; 15 - Burningham et al. 2011; 16 - Bowler et al. 2010; 17 - Scholz 2010; 18 - Burgasser et al. 2000a; 19 - Strauss et al. 1999; 20 - Burningham et al. 2009}
\tablenotetext{b}{Both components of the binary are in the GMOS field of view.}
\tablenotetext{c}{Scattered light from the primary prevented the determination of $i$.}
\tablecomments{$iz$ are on the GMOS-North and GMOS-South natural systems, see J{\o}rgensen (2009) and \S 2.2.  }
\end{deluxetable}
\clearpage

\begin{deluxetable}{lcrrrr}
\tabletypesize{\scriptsize}
\tablewidth{0pt}
\tablecaption{GMOS $i$ and $z$ Photometry for L and T  Dwarfs Transformed to the SDSS System}
\tablehead{
\colhead{Name} & \colhead{Spectral} & \colhead{$i$(err)} & \colhead{$z$(err)} & \colhead{$i$(err)}   & \colhead{$z$(err)}\\
\colhead{RA/Dec.} & \colhead{Type} &\colhead{} & \colhead{}  & \colhead{SDSS DR8} & \colhead{SDSS DR8}\\
}
\startdata
2MASSI J0415195$-$093506   & T8 & 23.45(0.09)  & 19.40(0.09)  & &   \\
SDSSp J042348.57$-$041403.5  & T0 & 20.14(0.08) &  17.31(0.08) &  20.19(0.04) & 17.29(0.01)  \\
2MASS J05591914$-$1404488  & T4.5 & 21.35(0.09) &  17.22(0.08)  & &  \\
UGPS J072227.51$-$054031.2  & T10 & 24.32(0.12)  & 20.52(0.10)  & &  \\
2MASS J07290002$-$3954043   & T8 &  24.18(0.13)  & 19.67(0.09) &   & \\
SDSS J075840.33$+$324723.4  & T2 & 21.86(0.09) &  18.07(0.08)  & 21.93(0.13) & 17.96(0.02)  \\
SDSS J083048.80$+$012831.1  & T4.5 & 24.13(0.16)  & 19.58(0.10)  &  &  19.40(0.07)\\
SDSSp J083717.22$-$000018.3  & T1 & 23.79(0.13)  & 20.02(0.11)  & 23.48(0.50) & 19.83(0.10) \\
ULAS J092624.76$+$071140.7   & T3.5 &   25.03(0.18) &  20.83(0.12)  &   &  \\
2MASSI J0937347$+$293142  &  T6 & 22.53(0.09)  & 18.01(0.08)  & &  \\
ULAS J095047.28$+$011734.30  & T8 & 25.91(0.21)  & 21.93(0.11)  &  & \\
2MASSI J1047538$+$212423 & T6.5 & 23.52(0.11)  & 18.95(0.09)  &   & \\
SDSS J104829.21$+$091937.8   &  T2.5 & 23.53(0.11)  & 19.66(0.10)  & 24.20(0.73) & 19.52(0.08)   \\
SDSSp J111010.01$+$011613.1  & T5.5 &  24.26(0.17) &  19.83(0.10)  & 23.92(0.78) & 19.67(0.10) \\
2MASS J11145133$-$2618235 & T7.5  &   23.69(0.13) &  19.49(0.10)  &  & \\
ULAS  J115759.04$+$092200.7  & T2.5 & 24.19(0.13) &  20.18(0.11)  &   & 19.92(0.15) \\
2MASS J12314753$+$0847331  & T5.5 & 22.75(0.10) &  18.74(0.10)  &  22.76(0.29) & 18.91(0.04)  \\
ULAS J130041.73$+$122114.7  &  T8.5 & 24.78(0.11)  & 20.29(0.09) & 23.28(0.57) & 20.04(0.16) \\
SDSS J141624.08$+$134826.7A & L7  & 18.39(0.08)  & 15.86(0.08) & 18.38(0.01) & 15.91(0.01) \\
SDSS J141623.94$+$134836.3B & T7.5 & 25.21(0.26)  & 20.87(0.09) & &    \\
Gliese 570D (2MASS J14571496$-$2121477)  & T7.5 & 23.08(0.09) &  18.90(0.09)  & & \\
SDSSp J162414.37$+$002915.6  & T6 & 22.95(0.09)  & 18.89(0.08) & 22.82(0.27) & 19.07(0.04) \\
SDSSp J175032.96$+$175903.0  &  T3.5 & 23.73(0.10)  & 19.46(0.09)  &  23.76(0.44) & 19.59(0.06) \\
Wolf 940B (ULAS J214638.83$-$001038.7)  & T8.5 &  &  21.88(0.14)  & & \\
2MASSI J2339101$+$135230  & T5 & 23.38(0.10) &  19.50(0.09) &  23.41(0.45) & 19.42(0.07)  \\
2MASSI J2356547$-$155310  & T5.5 & 22.92(0.10) &  19.27(0.09)  &   & \\
\enddata
\end{deluxetable}
\clearpage

\begin{deluxetable}{lll}
\tabletypesize{\scriptsize}
\tablewidth{0pt}
\tablecaption{Observational Properties of UGPS J072227.51$-$054031.2}
\tablehead{
\colhead{Measurement} & \colhead{Value}  & \colhead{Source} \\
}
\startdata
$i$ (AB SDSS) & 24.32 $\pm$ 0.12 & this work \\
$z$ (AB SDSS) & 20.52  $\pm$ 0.10 & this work \\
$Y$ (Vega MKO)  & 17.37   $\pm$ 0.02 & Lucas et al. 2010 \\
$J$ (Vega MKO)  & 16.52   $\pm$ 0.02 & Lucas et al. 2010 \\
$H$ (Vega MKO)  & 16.90   $\pm$ 0.02 & Lucas et al. 2010 \\
$K$ (Vega MKO)  & 17.07   $\pm$ 0.08 & Lucas et al. 2010 \\
$L^{\prime}$  (Vega MKO) & 13.4  $\pm$ 0.3 & Lucas et al. 2010 \\
W1(3.4) (Vega WISE)  & 15.15  $\pm$ 0.05 & Kirkpatrick et al. 2011 \\
3.6 (Vega IRAC)   & 14.28  $\pm$ 0.05 & Lucas et al. 2010 \\
4.5 (Vega IRAC)  & 12.19   $\pm$ 0.04 & Lucas et al. 2010 \\
W2(4.6)  (Vega WISE)  & 12.17  $\pm$ 0.03 & Kirkpatrick et al. 2011 \\
$N$  (Vega)  & 10.28   $\pm$ 0.24 & Lucas et al. 2010 \\
W3(12)  (Vega WISE)  & 10.18  $\pm$ 0.06 & Kirkpatrick et al. 2011 \\
W4(22)  (Vega WISE)  & $>$8.64  & Kirkpatrick et al. 2011 \\
$v_{\rm rotational}$ (km s$^{-1}$)  & 40 $\pm$ 10 & Bochanski et al. 2011 \\ 
$v_{\rm radial}$ (km s$^{-1}$)  & 46.9 $\pm$ 2.5 & Bochanski et al. 2011 \\
$\mu_{\rm RA}$ (mas yr$^{-1}$)  & $-904.14$ $\pm$  1.71 & this work \\
$\mu_{\rm dec}$ (mas yr$^{-1}$)  & 352.025 $\pm$  1.21 & this work \\
${\pi}$ (mas)  & 242.8 $\pm$  2.40 &  this work\\
$U$ (km s$^{-1}$){\tablenotemark{a}}  & 37 $\pm$  2  & this work, toward the Galactic anticenter\\
$V$ (km s$^{-1}$){\tablenotemark{a}}  & $-8 \pm$ 1 & this work, in the direction of Galactic rotation\\
$W$ (km s$^{-1}$){\tablenotemark{a}}  & $-2.4 \pm$ 0.3 & this work, toward the North Galactic Pole\\
\enddata
\tablenotetext{a}{$UVW$ are corrected to the local standard of rest using the solar values from Co\c{s}kuno\u{g}lu et al. (2011): $U_{\odot}=-8.50$, $V_{\odot}=13.38$ and $W_{\odot}=6.49$~km s$^{-1}$. The velocities have been calculated using the updated parallax and proper motion presented in this work, and the radial velocity of Bochanski et al. (2011).} 
\end{deluxetable}
\clearpage

\begin{deluxetable}{rrrcrc}
\tablewidth{0pt}
\tablecaption{Range of Physical Properties for UGPS J072227.51$-$054031.2}
\tablehead{
\colhead{$T_{\rm eff}$} & \colhead{$\log g$} & \colhead{$\log L/L_{\odot}$} & \colhead{Mass} & \colhead{Radius}   & \colhead{Age}\\
\colhead{K} & \colhead{cm s$^{-2}$} &\colhead{} & \colhead{$M_{\rm Jupiter}$}  & \colhead{$R_{\odot}$} & \colhead{Gyr}\\
}
\startdata
492  &  3.52 & $-$6.05 & 2.1 & 0.128 & 0.04 \\
518  &  4.48 & $-$6.13 & 13.2 & 0.108 & 1.4  \\
550  &  5.00 & $-$6.17 & 31.3 & 0.091 & 6.6 \\

\enddata
\end{deluxetable}

\end{document}